\documentclass[twocolumn,secnumarabic,amssymb, nobibnotes, aps, prb,groupedaddress,superscriptaddress]{revtex4-2}
\usepackage{amsmath,graphicx,latexsym,times,color}
\usepackage{setspace}
\usepackage{hyperref}
\usepackage{array}
\usepackage{titlesec}
\usepackage{physics}
\usepackage{gensymb}
\usepackage{nccmath}
\usepackage{empheq} %% loads mathtools, which loads amsmath

\newcommand{\V}[1]{\ensuremath{\mathbf{#1}}} %Vector

\let\oldtimes\times  % Make the times "x" use less spacing
\renewcommand\times{{\oldtimes}}

\newcommand{\parallelsum}{\mathbin{\!/\mkern-5mu/\!}}
	
\renewcommand{\vec}[1]{\mathbf{#1}}

\begin{document}
	
\title{Tuning the magnetic interactions in van der Waals Fe$_3$GeTe$_2$ heterostructures: \\
A comparative study of \textit{ab initio} methods}
	
\author{Dongzhe Li}
\email{dongzhe.li@cemes.fr}
\affiliation{CEMES, Universit\'e de Toulouse, CNRS, 29 rue Jeanne Marvig, F-31055 Toulouse, France}
	
\author{Soumyajyoti Haldar}
\affiliation{Institute of Theoretical Physics and Astrophysics, University of Kiel, Leibnizstrasse 15, 24098 Kiel, Germany}

\author{Tim Drevelow}
\affiliation{Institute of Theoretical Physics and Astrophysics, University of Kiel, Leibnizstrasse 15, 24098 Kiel, Germany}

\author{Stefan Heinze}
\affiliation{Institute of Theoretical Physics and Astrophysics, University of Kiel, Leibnizstrasse 15, 24098 Kiel, Germany}
\affiliation{Kiel Nano, Surface, and Interface Science (KiNSIS), University of Kiel, Germany}

\date{\today}
	
\begin{abstract}
	We investigate the impact of mechanical strain, stacking order, and external electric fields on the magnetic interactions of a two-dimensional (2D) van der Waals (vdW) heterostructure in which a 2D ferromagnetic metallic Fe$_3$GeTe$_2$ monolayer is deposited on germanene. Three distinct computational approaches based on \textit{ab initio} methods are used, and a careful comparison is given: (i) The Green's function method, (ii) the generalized Bloch theorem, and (iii) the supercell approach. First, the shell-resolved exchange constants are calculated for the three Fe atoms within the unit cell of the freestanding Fe$_3$GeTe$_2$ monolayer. We find that the results between methods (i) and (ii) are in good qualitative agreement and also with previously reported values. An electric field of ${\mathcal E}= \pm 0.5$~V/{\AA} applied perpendicular to the Fe$_3$GeTe$_2$/germanene heterostructure leads to significant changes of the exchange constants. We show that the Dzyaloshinskii-Moriya interaction (DMI) in Fe$_3$GeTe$_2$/germanene is mainly dominated by the nearest neighbors, resulting in a good quantitative agreement between methods (i) and (ii). 
 Furthermore, we demonstrate that the DMI is highly tunable by strain, stacking, and electric field, leading to a large DMI comparable to that of ferromagnetic/heavy metal (FM/HM) interfaces, which have been recognized as prototypical multilayer systems to host isolated skyrmions. The geometrical change and hybridization effect explain the origin of the high tunability of the DMI at the interface. The electric-field driven DMI obtained by method (iii) is in qualitative agreement with the more accurate \textit{ab initio} method used in approach (ii). However, the field-effect on the DMI is overestimated by method (iii) by about 50\%. This discrepancy is attributed to the different implementations of the electric field and basis sets used in the \textit{ab initio} methods applied in (ii) and (iii). The magnetocrystalline anisotropy energy (MAE) can also be drastically changed by the application of compressive or tensile strain in the Fe$_3$GeTe$_2$/germanene heterostructure. The application of an electric field, in contrast, leads only to relatively small changes of the MAE for electric fields of up to 1~V/{\AA}.
\end{abstract}

\maketitle

\section{Introduction}
\label{intro}

Magnetic skyrmions~\cite{Bogdanov1989} -- topologically protected chiral spin structures with particle-like properties -- have attracted tremendous attention due to their potential application in next-generation spintronics devices such as racetrack memories \cite{fert2013}, logic gates \cite{Luo2018}, artificial synapses for neuromorphic computing \cite{song2020skyrmion}, and qubits for quantum computing \cite{Psaroudaki2021}. The formation of magnetic skyrmions is due to the competition between the Heisenberg exchange and Dzyaloshinskii-Moriya interaction (DMI) \cite{Moriya1960,Dzyaloshinskii1957,Fert1980,bode2007chiral}
together with the magnetocrystalline anisotropy energy (MAE). In particular, the DMI
plays an essential role in stabilizing skyrmions since it favors canted spin configurations with a unique rotational sense. The DMI results from spin-orbit coupling and is only non-zero for systems with
broken inversion symmetry. Magnetic skyrmion lattices were discovered in experiments on bulk chiral
magnets~\cite{Muhlbauer2009skyrmion,Yu2010} and subsequently in epitaxial ultrathin films~\cite{heinze2011spontaneous}. Isolated magnetic skyrmions were observed in 
ultrathin transition-metal films at low temperatures 
\cite{Romming2013,herve2018,meyer2019isolated} and at 
room temperature in magnetic multilayers \cite{moreau2016additive,boulle2016room,woo2016observation,Soumyanarayanan_2017},
in ferrimagnets \cite{Caretta2018}, and in synthetic antiferromagnets \cite{legrand2020room}.

Recently, long-range magnetism was reported in two-dimensional (2D) materials \cite{gong2017discovery,huang2017layer,deng2018gate}. This provides a promising alternative avenue for exploring topological spin structures in atomically thin layers. Several recent experiments reported the observation of skyrmions in 2D vdW heterostructures, such as at an Fe$_3$GeTe$_2$/WTe$_2$ interface \cite{wu2020neel}, in a Fe$_3$GeTe$_2$/Co/Pd multilayer \cite{yang2020creation}, and at an Cr$_2$Ge$_2$Te$_6$/Fe$_3$GeTe$_2$ interface \cite{wu2021van}. Moreover, magnetic domain walls \cite{Yang_2022} and nonreciprocal magnons \cite{Costa2020} were reported in the Fe$_3$GeTe$_2$ surface. The origin of skyrmions in these systems was attributed to the interfacial DMI. A comprehensive material survey has been done by \textit{ab initio} calculations to explore the DMI in 2D magnets. The family of monolayer Janus vdW magnets has been predicted to possess large enough DMI to allow stable skyrmions \cite{Liang2020,Yuan2020,Changsong2020,Cui2020strain,jiang2021topological,shen2022strain}. 
N\'eel-type magnetic skyrmions were also observed in Fe$_3$GeTe$_2$ crystals, and attributed to the DMI due to oxidized interfaces \cite{Park2021}. In addition, it has been proposed that skyrmions can be stabilized in 2D vdW multiferroic heterostructures \cite{sun2020controlling}, Moir\'e of vdW 2D magnets \cite{tong2018skyrmions}, and even in centrosymmetric materials \cite{amoroso2020spontaneous} induced by exchange frustration. For most 2D magnets, the DMI is absent due to inversion symmetry. It is possible to break the inversion 
symmetry by designing various 2D vdW heterostuctures and applying an electric field, or strain \cite{gong2019two,jiang2021recent}. This indicates the possibility of tuning DMI via external stimuli in 2D vdW heterostuctures.

%\textcolor{red}{I guess one would not have said that the DMI calculation is so straight forward about 10 years ago. Nowadays, it is of course possible by different codes. Maybe we should also cite work by the KKR method, e.g. from the Budapest group of Szunyogh.}
From the theoretical point of view, the calculation of the DMI at the \textit{ab initio} level is, in principle, relatively straightforward, nevertheless, complications can arise in practice. 
Several approaches have been introduced based on different first-principles
methods and used by numerous groups to calculate the DMI for various material 
classes \cite{bode2007chiral,Ferriani2008,Heide2009,Yanes2013,Kashid2014,Zimmermann2014,dupe2014tailoring,Simon2014,Yang2015,Vida2016,Yamamoto2017,Simon2018,Jadaun2020,Liang2020}.
%\textcolor{blue}{Please add a few more citations here 
%incl. the Budapest group and some for 2D materials.} \textcolor{red}{I have added two papers from the Budapest group of Szunyogh}
Unfortunately, mostly 
%SH B. Zimmermann et al. provided a check between different approaches
without a sufficient cross-check between them. This invites a detailed benchmark study to validate different approaches, however, so far this has only been performed for the ultrathin film system of a Co monolayer on Pt(111) \cite{Zimmermann2019}. Such comparative studies are crucial to understanding the origin of skyrmion stability, particularly for the newly discovered 2D magnets.

Here, using \textit{ab initio} calculations, we compare systematically three current state-of-the-art approaches to extract magnetic interaction parameters in Fe$_3$GeTe$_2$ (FGT) based heterostructures, namely (i) by the Green's function method~\cite{liechtenstein1987local}, (ii) by using the generalized Bloch 
theorem (gBT)~\cite{Sandratskii1986,Kurz2004,Heide2009,Zimmermann2019}, or (iii) by using the 
supercell approach~\cite{Yang2015}. 
%Calculations using approaches (i) and (iii) were performed with the QuantumATK code which is
%based on the linear combination of atomic orbitals, while calculations using approach (ii)
%were carried out using the full-potential linearized augmented plane wave method as implemented in the %FLEUR code.
First we study the shell-resolved exchange interaction between the Fe atoms of the different layers in 
a freestanding FGT monolayer. We find that the approaches (i) and (ii) are in good qualitative agreement. 
We then focus on the structural and magnetic properties of the 2D vdW heterostructure of an FGT monolayer deposited on germanene
%SH I think we should specify the heterostructure here, FGT heterostructures 
under strain, stacking, and electric field.
We find that a small compressive strain ($\gamma$) of about 3\% can significantly enhance the DMI in FGT 
heterostructures by more than 400\% compared to the value
%than the one 
without strain. The variation of DMI is mainly due to the geometrical change of the FGT monolayer. Such a large DMI is comparable to that in state-of-the-art FM/HM heterostructures, which have been demonstrated as prototypical multilayer systems to host individual skyrmions even at room temperature. Furthermore, the DMI can be substantially modified via different stacking geometry due to the hybridization effect at the interface. 

Upon applying an electric field the strength of the DMI varies almost linearly and can even change sign when a strong electric field (${\cal E}>$ 1 V/\AA) is applied. The exchange constants are also considerably modified due to an electric field while the effect on the magnetocrystalline anisotropy energy (MAE) is small. However, the MAE is dramatically reduced to 25\% of its original value at a compressive strain of $\gamma=-3\%$. In connection with the exchange frustration in FGT/germanene these large changes of DMI and MAE open the possibility of zero-field magnetic skyrmions \cite{Dongzhe2022_fgt}. For the DMI in FGT/Ge, the three theoretical approaches are in good qualitative agreement, and we also discuss quantitative comparison in detail.

The paper is organized as follows. In Sec.~\ref{methods}, we describe the three theoretical approaches used to obtain the relevant spin-spin interaction parameters by mapping the \textit{ab initio} DFT calculations onto an extended Heisenberg model. In Sec.~\ref{results}, we examine the Heisenberg exchange for free-standing FGT followed by the DMI and MAE for FGT heterostuctures. Different theoretical approaches are carefully benchmarked. We further investigate the effects of biaxial strain, stacking configuration as well electric field on the magnetic interactions in FGT heterostuctures. Finally, we summarize our main conclusions in Sec.~\ref{concl}.

\section{Methods and computational details}
\label{methods}

In order to describe the magnetic properties of FGT heterostructures, we use the extended Heisenberg model for the spins of Fe atoms in the hexagonal structure:

\begin{equation}\label{model}
\begin{split}
H & =-\sum_{ij}J_{ij}(\vec{m}_i \cdot \vec{m}_j)-\sum_{ij}\vec{D}_{ij} \cdot(\vec{m}_i \times \vec{m}_j)\\
&-\sum_i K_i (m_i^z)^2 
\end{split}
\end{equation}

where $\V{m}_i$ and $\V{m}_j$ are normalized magnetic moments at positions $\vec{R}_i$ and $\vec{R}_i$ respectively. The three magnetic interaction terms correspond to the Heisenberg isotropic exchange, the DMI, and the MAE, respectively, and they are characterized by the parameters $J_{ij}$, $\vec{D}_{ij}$, and $K_i$ in the related terms. Note, that by using Eq.~(\ref{model}) it is assumed that the magnetic moments are constant.

During the past decade, in order to obtain the parameters very accurately in Eq.~(\ref{model}), several approaches based on density functional theory (DFT) have been developed, which is frequently named \textit{ab initio} atomistic spin model. In this work, we apply three different approaches for the calculation of magnetic interactions in FGT vdW heterostructures: (i) The Green's function method \cite{liechtenstein1987local,Szilva2013} (also known as the Liechtenstein formula) employing infinitesimal rotations; (ii) The generalized Bloch theorem (gBT) \cite{Sandratskii1986} which allows calculating the total energy of
%simulating 
spin-spirals of any wave vector $\vec{q}$ in magnetic nanostructures~\cite{Kurz2004}; (iii) The supercell approach \cite{Yang2015} which is straightforward but computationally heavy due to the comparison of total energies in a supercell geometry. We performed DFT calculations using two community \textit{ab initio} codes which differ in their choice of basis set: The {\tt QuantumATK (QATK)} package \cite{smidstrup2019} uses an expansion of electronic states in a linear combination of atomic orbitals (LCAO) while the {\tt FLEUR} code \cite{fleur_v26} is based on the full-potential linearized augmented plane wave (FLAPW) formalism. The former is computationally very efficient, while the latter ranks amongst the most accurate implementations of DFT. In the following, we denote the three different approaches as LCAO-Green, FLAPW-gBT, and LCAO-supercell for simplicity. Apart from the methods presented above, there are also other approaches widely used in the community for calculations of spin-spin interactions, e.g., the four-state method \cite{xiang2013magnetic} and the machine learning approach \cite{Hongyu2022}.

\subsection{The Green's function method: LCAO-Green}
\label{Green_approach}
Throughout this paper, vectors are denoted with bold characters while matrices are represented by bold plus single underline (e.g., $\underline{\vec{G}_{ij}}$). Moreover, $L$ represent orbital index while $\boldsymbol{\sigma} = (\sigma_x,\sigma_y,\sigma_z) $ is spin index.

The variation of total energy due to the spin interactions in Eq. \ref{model}, we
obtain the following variation with respect to the $\vec{m}_i$ and $\vec{m}_j$:

\begin{equation}\label{green0}
\begin{split}
\delta E_{ij} =& -2J_{ij}(\delta{\vec{m}_i} \cdot \delta{\vec{m}_j}) \\
& -2\delta{\vec{m}_i}\mathbf{\underline{J}}_{ij}^{\text{ani}} \delta{\vec{m}_j} \\
& -2\vec{D}_{ij}\cdot(\delta{\vec{m}_i} \times \delta{\vec{m}_j})
\end{split}
\end{equation}

where the first term represents the isotropic exchange (i.e., Heinsenberg), the second term is the the symmetric anisotropic exchange, where $\mathbf{\underline{J}}_{ij}^{\text{ani}}$ is a $3 \times 3$ symmetric tensor. The last term corresponds to the DMI,

Green's function method treats the local rigid spin rotation as a perturbation. Using the force theorem, the total energy variation due to the two-spin interaction between sites $i$ and $j$ is 

\begin{equation}\label{green1}
\delta E_{ij}=-\frac{2}{\pi}\int_{-\infty}^{E_{\text{F}}} dE \Im \Tr[\delta\mathbf{\underline{H}}_{ii}\mathbf{\underline{G}}_{ij}\delta\mathbf{\underline{H}}_{jj}\mathbf{\underline{G}}_{ji}]
\end{equation}
where $\mathbf{\underline{H}}_{ii} = \delta \vec{e}_i \cdot \boldsymbol{\sigma}$ and  $\mathbf{\underline{G}}_{ij} = G_{ij}^0 \mathbf{\underline{I}} + \mathbf{G}_{ij} \cdot \boldsymbol{\sigma}$ are real-space Hamiltonian and Green's function. Here, $\vec{e}_i = \vec{m}_i$ is a unit orientation vector (normalized to 1).

Then, in Eq. \ref{green1}, if we take trace in both orbital ($L$) and spin space ($\boldsymbol{\sigma}$), we end up with the following expression:

\begin{equation}\label{green1_2}
\footnotesize 
\begin{split}
 \Tr_{L,\boldsymbol{\sigma}}[\delta\mathbf{\underline{H}}_{ii}\mathbf{\underline{G}}_{ij}\delta\mathbf{\underline{H}}_{jj}\mathbf{\underline{G}}_{ji}]  =& -2 [G_{ij}^0G_{ji}^0  -\sum_{\mu \in (x,y,z)}\mathbf{\underline{G}}_{ij}^{\mu} \mathbf{\underline{G}}_{ji}^{\mu}]\delta \vec{e}_i \delta \vec{e}_j \\
  & -2 \sum_{\mu,\nu \in (x,y,z)}\delta \vec{e}_i^{\mu}(\mathbf{\underline{G}}_{ii}^{\mu} \mathbf{\underline{G}}_{ji}^{\nu}+\mathbf{\underline{G}}_{ij}^{\mu} \mathbf{\underline{G}}_{ji}^{\nu})\delta \vec{e}_j^{\nu} \\
  & -2 \vec{D}_{ij} \cdot (\delta \vec{e}_i \times \delta \vec{e}_j)]
\end{split}
\end{equation}

To simplify for the notation of Eq. \ref{green1_2}, we define a central quantity for the Green function method, namely the $\underline{\mathbf{A}}$ matrix, which has a $4 \times 4$ size as follows.

\begin{equation}\label{green2}
{A}_{ij}^{\mu\nu}=-\frac{1}{4\pi}\int_{-\infty}^{E_{\text{F}}}dE\Tr_{L}[\underline{\mathbf{G}}_{ij}^{\mu}\underline{\mathbf{G}}_{ji}^{\nu}]
\end{equation}
where indices $\mu$ and $\nu$ run over 0, $x$, $y$, or $z$.

Finally, comparing Eq. \ref{green1_2} to Eq. \ref{green0}, the Heisenberg exchange and the DMI can be expressed by using only the $\mathbf{A}$ matrix as follows.

\begin{equation}\label{green3}
J_{ij}=2\Im({A}_{ij}^{00}-{A}_{ij}^{xx}-{A}_{ij}^{yy}-{A}_{ij}^{zz})\frac{\underline{\mathbf{\Delta}}_{ii}\underline{\mathbf{\Delta}}_{jj}}{4}
\end{equation}

\begin{equation}\label{green3-2}
J_{ij}^{\text{ani}}=2\Im({A}_{ij}^{\mu \nu}+{A}_{ij}^{\nu \mu})\frac{\underline{\mathbf{\Delta}}_{ii}\underline{\mathbf{\Delta}}_{jj}}{4}
\end{equation}

\begin{equation}\label{green4}
\mathbf{D}_{ij}^{\mu}=2\Re({A}_{ij}^{0\mu}-{A}_{ij}^{\mu 0})\frac{\underline{\mathbf{\Delta}}_{ii}\underline{\mathbf{\Delta}}_{jj}}{4}
\end{equation}

where $\underline{\mathbf{\Delta}}_{ii}=(\underline{\mathbf{H}}_{ii}^{\uparrow}-\underline{\mathbf{H}}_{ii}^{\downarrow})$ is the on-site difference between the spin-up and -down part of the Hamiltonian matrix. 

If we neglect SOC, the DMI vanishes, $\underline{\mathbf{G}}^{x} = \underline{\mathbf{G}}^{y}=0$, $\underline{\mathbf{G}}^{0}=1/2(\underline{\mathbf{G}}^{\uparrow}+\underline{\mathbf{G}}^{\downarrow})$, $\underline{\mathbf{G}}^{z}=1/2(\underline{\mathbf{G}}^{\uparrow}-\underline{\mathbf{G}}^{\downarrow})$, we arrive at the original Liechtenstein-Katsnelson-Antropov-Gubanov (LKAG) formula \cite{liechtenstein1987local} which was proposed in 1987,

\begin{equation}\label{green5}
J_{ij}=-\frac{1}{4\pi}\int_{-\infty}^{E_{\text{F}}}dE\Im \Tr[\underline{\mathbf{\Delta}}_{ii}\underline{\mathbf{G}}_{ij}^{\uparrow}\underline{\mathbf{\Delta}}_{jj}\underline{\mathbf{G}}_{ji}^{\downarrow}]
\end{equation}

where $\underline{\mathbf{G}}_{ij}$ becomes hermitian.

Note that the derivations above are general for orthogonal and non-orthogonal basis sets. Please refer to Ref. \cite{Oroszl2019} for a detailed demonstration within the non-orthogonal basis set.

Our Green's function calculations were performed using {\tt QATK} \cite{smidstrup2019} in two steps: (i) We performed LCAO-DFT calculations with SOC in order to construct the tight-binding like Hamiltonian matrix $\mathbf{H}_{ij}$ and the overlap matrix $\mathbf{S}_{ij}$. (ii) The magnetic exchange parameters were evaluated as described above by Eqs.~(\ref{green1}-\ref{green5}). For LCAO-DFT calculations on FGT monolayers and on FGT/Ge heterostructures, the energy cutoff for the density grid sampling was set to 150 Hartree, and a $28 \times 28$ $\vec{k}$-point mesh was adopted for the Brillouin zone (BZ) integration. For magnetic exchange calculations, we used a much denser $\vec{k}$-point mesh of $48 \times 48$, 60 circle contour points, and 13$^{\text{th}}$ nearest neighbors (NN) in order to obtain accurate numerical integration. Using these parameters, we extracted $J_{ij}$ and $\vec{D}_{ij}$ parameters with an accuracy of 0.01 meV.  Note that this approach, i.e., infinitesimal spin rotations, fits well to magnetic skyrmions in which we often have large noncollinear spin structures.

\subsection{The generalized Bloch theorem: FLAPW-gBT}
\label{gBT_approach}

The second approach employs the FLAPW method as implemented in the {\tt FLEUR} code \cite{fleur_v26} and is based on the generalized Bloch theorem (gBT) \cite{Sandratskii1986,Kurz2004}. It allows considering spin-spirals of any wave vector $\vec{q}$ for systems without SOC. We first self-consistently compute 
within the scalar-relativistic approximation
the energy dispersion, $E_{\text{ss}}(\vec{q})$, of homogeneous flat spin spirals \cite{Kurz2004}
which are characterized by a wave vector $\V{q}$ and an angle $\phi= \V{q} \cdot \V{R}$ between adjacent magnetic moments separated by lattice vector $\V{R}$. 

As a second step, the DMI is computed within first-order perturbation theory on the self-consistent spin spiral state \cite{Heide2009,Zimmermann2014,Zimmermann2019}. 
The energy variation $\delta \epsilon_{\vec{k},\nu} (\mathbf{q})$ of these states due to the SOC Hamiltonian can be written as
\begin{equation}\label{gBT1}
\delta \epsilon_{\vec{k},\nu} (\mathbf{q}) = \bra{\Psi_{\vec{k},\nu}(\vec{q})}H_{\text{SOC}}\ket{\Psi_{\vec{k},\nu}(\vec{q})},
\end{equation}
where $\ket{\Psi_{\vec{k},\nu}(\vec{q})}$ are the self-consistent solutions
in the scalar-relativistic approximation, $\mathbf{k}$ is the Bloch vector,
and $\nu$ is the band index.
%runs over all occupied states.
By integration over the Brillouin zone and summation over all occupied
bands $\nu$ this gives the total energy contribution 
for spin spirals due to SOC denoted as $E_{\rm DMI}(\mathbf{q})$.

%The SOC contribution at each binding energy is given by
%\begin{equation}\label{gBT2}
%\epsilon_{\text{SOC}}(E,\vec{q})=\sum_{\nu}\int \delta \epsilon_{\vec{k},\nu}\delta(E-E_{\vec{k}, \nu})d\vec{k}.
%\end{equation}
%After that, the DMI energy is defined as
%\begin{equation}\label{gBT22}
%E_{\text{DMI}}(\vec{q})=\int^{E_F} \epsilon_{\text{SOC}}(E',\vec{q})dE'
%\end{equation}

We map the energy dispersion in the scalar-relativistic approximation,
$E_{\text{SS}}(\mathbf{q})$, and the energy contribution to spin spirals
due to DMI, $E_{\text{DMI}}(\mathbf{q})$, 
to the atomistic spin model, Eq.~(\ref{model}),
%with a set of fitting functions (e.g., cosine or sine), 
in order to extract the exchange constants, $J_{ij}$, and the
magnitudes of the $\vec{D}_{ij}$, respectively. One of the key advantages of 
the gBT approach is that even incommensurate spin spirals and those with a 
large $\vec{q}$ can be treated very efficiently 
in the chemical unit cell, i.e., without the need for large supercells. 

\begin{figure}[t]%---------------------------------------------------------------------------------------------------------------------Fig_1
\centering
\includegraphics[width=1\linewidth]{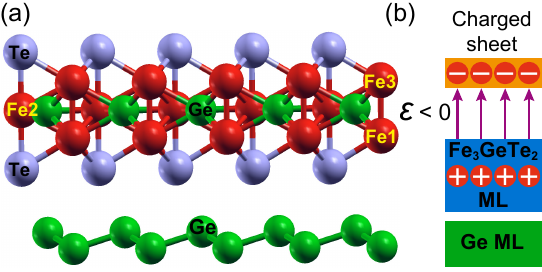}
\caption{\label{FGT_ge_structure} (a) Sketch of the studied 2D magnetic
vdW heterostructure of an Fe$_3$GeTe$_2$ monolayer (ML) on germanene. 
%exposed to an out-of-plane uniform electric field. 
Three nonequivalent Fe atoms are indicated as red spheres: Fe1 (bottom, interface atom), Fe2 (middle), and Fe3 (top). Te and Ge atoms are given by grey and green spheres, respectively.
(b) The electric field is created by a charged sheet located at 5.3 \AA~above the FGT. The direction of electric field lines for ${\cal E}<0$ is shown.}
\end{figure}

We used a cutoff parameter for the FLAPW basis functions of $k_{\text{max}}$ = 4.0  a.u.$^{-1}$, and we included basis functions including spherical harmonics up to $l_{\text{max}}$ = 8. The muffin tin radii used for Fe, Ge, and Te are 2.10 a.u., 2.10 a.u., and 2.63 a.u., respectively. Moreover, we treated 3$s$, 3$p$ and 4$d$ states by local orbitals for Fe and Te, respectively. To extract the $J_{ij}$ and $\vec{D}_{ij}$ parameters, we converge the total energy of flat spin-spiral states (without SOC and with one-shot SOC) using a $33 \times 33$ $\vec{k}$-point mesh. For conical spin spiral calculations, we increased the $\V{k}$-point mesh up to $49 \times 49$ since the energy dispersion amplitude is much smaller than for the flat ones. We model the effect of a uniform electric field by placing a charged sheet in the vacuum region of FGT/Ge (see Fig.~\ref{FGT_ge_structure}),
i.e.~using the same methodology and sign convention of the electric field as in Ref.~\onlinecite{paul2022electric}. We maintain the charge neutrality of the whole system by adding or removing the same amount of opposite charge to or from the interface. Finally, we computed the MAE using the force theorem using a denser $\V{k}$-mesh of 64 $\times$ 64.

%-------------------------------------------------------------------------------------------------------------------------------

\subsection{The supercell approach: LCAO-supercell}
\label{supercell_approach}

%\textcolor{red}{If one uses the 90 degree spin spiral in the 4 atom unit cell, the super cell approach leads to the nearest-neighbor DMI constant. The effective D is obtained in the limit of small q, i.e. large spin spiral periods which is not reached in the super cell approach.} \textcolor{blue}{Yes, I agree with you.}
%In this approach, the effective DMI within the nearest-neighbor approximation, $D_{\text{eff}}$, for a specific system can be given by the energy difference between clockwise (CW) and counterclockwise (CCW) spin configurations. $d_{\parallelsum}$ can be obtained by the following formula \cite{Yang2015}:
In this approach, the DMI within the nearest-neighbor approximation for a specific system 
is calculated from
%can be given by 
the energy difference between a clockwise (CW) and a counterclockwise (CCW) 90$^\circ$
spin spiral calculated within a supercell.
%spin configurations. If one uses a supercell with a 90$^\circ$ spin spiral
$d_{\parallelsum}$ can be obtained by the following formula \cite{Yang2015}:
%\textcolor{red}{This formula applies to the 90 degree spiral, right?} \textcolor{blue}{Yes!}

\begin{equation}\label{supercell}
d_{\parallelsum}=(E_{\text{CCW}}-E_{\text{CW}})/8\sqrt{3}
\end{equation}

The corresponding micromagnetic DMI coefficient, $D$,
%SH $D_{\text{mm}}$, can be calculated as follows

\begin{equation}\label{supercell_M}
D=\frac{3\sqrt2d_{\parallelsum}}{N_{\text{F}}a^2}
\end{equation}
where $a$ and $N_{\text{F}}$ are the lattice constant and number of ferromagnetic layers, respectively.

Note, that by varying the supercell size one can go beyond nearest-neighbor DMI approximation.
However, the computational effort is much larger than using the spin spiral approach sketched above.

We used the {\tt QATK} code \cite{smidstrup2019} and a $4 \times 1$ supercell where first neighbor spins rotated as 90$^{\circ}$. 
A $7 \times 28$ $\vec{k}$-point mesh was adopted 
for the BZ integration. We included the effect of SOC self-consistently. Although the supercell 
approach is straightforward and SOC can be treated self-consistently, it is only limited to very 
large wave vectors.

\section{Results and discussions}
\label{results}

Throughout the paper, we use the following conventions. We use a minus sign in the Heisenberg exchange terms (cf.~Eq.~(1)), meaning that $J_{ij}>0$ ($J_{ij}<0$) favors ferromagnetic (antiferromagnetic) alignment
between moments $\vec{m}_i$ and $\vec{m}_j$. 
For the DMI constants, a positive (negative) sign
%SH: I think that we should only refer to the constants not
% to the DMI vectors, $\vec{D}_{ij}>0$ ($\vec{D}_{ij}<0$) 
denotes a preferred CW (CCW) rotational sense. 
In addition, without specification, we give in our work the
in-plane component of the DMI since the out-of-plane component has shown to be negligible forming skyrmions in 2D magnets \cite{Liang2020,du2022spontaneous}.
All effective quantities, such as the Heisenberg exchange, the DMI, and the MAE, are measured in meV/unit cell. Note that there are 3 Fe atoms per unit cell of FGT (Fig.~\ref{FGT_ge_structure}).

\subsection{Geometric properties}

We consider FGT heterostructures in which an FGT monolayer is deposited on germanene (denoted as FGT/Ge in the following). As shown in Fig.~\ref{FGT_ge_structure}(a), FGT adopts the space group (194) P6$_3$/\textit{mmc} and can be seen from the perspective of the atomistic spin model as a stack of three Fe hexagonal layers in hcp stacking. In the following, the top, center, and bottom (interface) atoms are denoted as Fe3, Fe2, and Fe1, respectively. 

The motivation behind the use of germanene as a nonmagnetic layer is as follows. We recently demonstrated that the buckled structure of germanene could enhance the structural asymmetry of FGT under strain \cite{Dongzhe2022_fgt}. More importantly, such an efficient strain-driven DMI control is general for FGT heterostructures with buckled substrates (e.g., silicene, antimonene). On the other hand, in experiments, FGT/graphene \cite{JLopes_2021}, and FGT/hBN \cite{wang2018tunneling} have been experimentally synthesized. Since germanene is very similar to graphene in many aspects, the FGT/Ge interface is expected to be feasible.

We have used the {\tt QATK} code with plane-wave basis sets for the atomic relaxation.
We employed the generalized gradient approximation (GGA), obtaining a relaxed lattice constant of 4.00~\AA~for the Fe$_3$GeTe$_2$ monolayer, which is in good agreement with experimental data ranging between about 
3.991 $\sim$ 4.03 {\AA}~\cite{Deiseroth2006,chen2013magnetic}. Then, germanene is matched at the interface with FGT with a lattice mismatch smaller than 1\%. The structures were fully relaxed until the energy and the forces on each atom were less than 10$^{-8}$ Ry and 10$^{-4}$ Ry/Bohr, respectively. We also took into account van der Waals interactions using semi-empirical dispersion corrections as formulated by Grimme \cite{grimme2010consistent}. We used the local density approximation (LDA) for the magnetic calculations with LCAO basis sets. We did not take into account Hubbard $U$ correction since LDA yields a magnetic moment of 1.76 $\mu_{\text{B}}$/Fe that compares well with experiments, as previously pointed out in Ref.~\onlinecite{Zhuang2016,deng2018gate}. It has been shown recently by S. Ghosh \textit{et al.} that using DFT+$U$ leads to a less favorable description of the magnetic properties for the FGT family \cite{ghosh2022unraveling}. The lowest-energy stacking configuration is the one where the Te atom is right above the center of the hexagonal ring of germanene with an optimized vdW gap of about 2.86~\AA~(see Fig.~\ref{FGT_ge_structure}b), which agrees well with previous results \cite{Junjie2021}.

\subsection{Free-standing FGT monolayer: Heisenberg exchange}
\label{free-layer}

	\begin{figure}[t]%---------------------------------------------------------------------------------------------------------------------Fig_2
	\centering
	\includegraphics[width=1.0\linewidth]{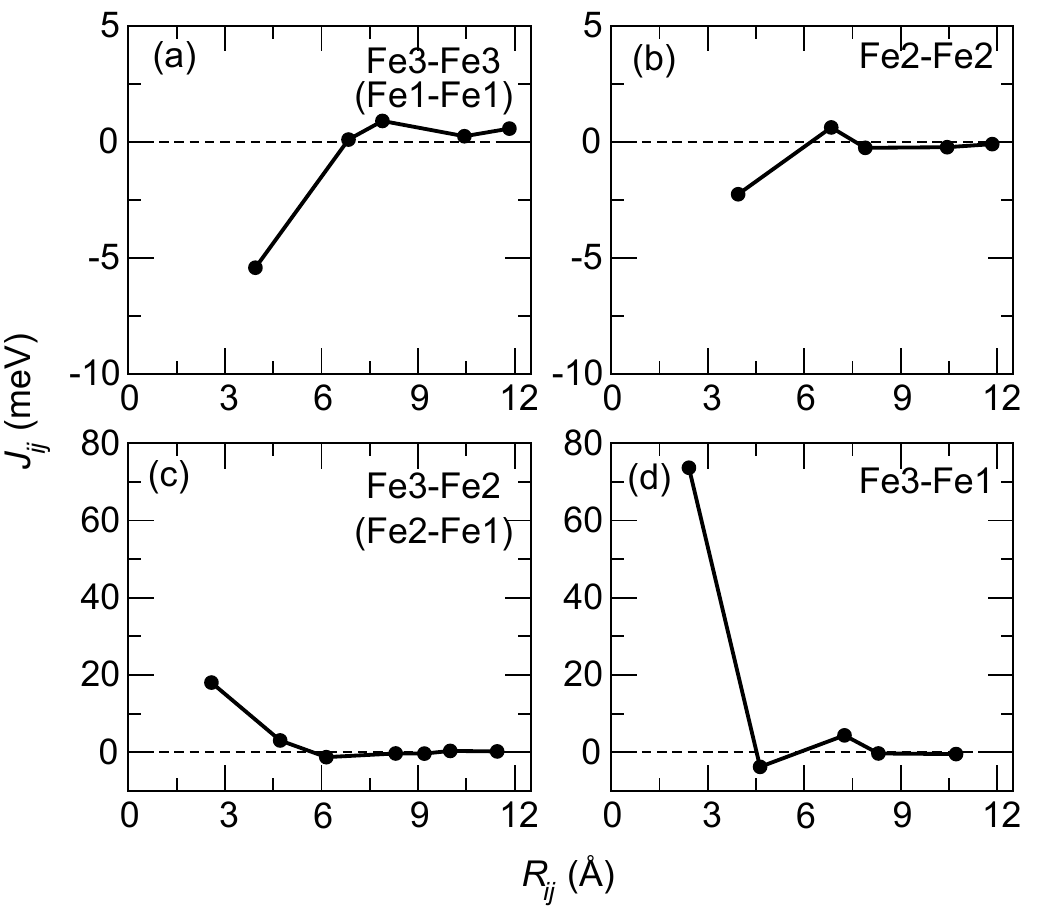}
	\caption{\label{Heisenberg_Green} Calculated Heisenberg exchange constants as a function of distance 
    $R_{ij}$ between sites $i$ and $j$
    for a free-standing FGT monolayer
     %. The results have been 
    obtained within the LCAO-Green approach. (a,b) Intralayer exchange interactions for Fe3-Fe3 (Fe1-Fe1) and Fe2-Fe2 spin pairs, respectively. (c,d) Interlayer exchange interactions for Fe3-Fe2 (Fe2-Fe1) and
    Fe3-Fe1 spin pairs, respectively. Note, that $J_{ij}>0$ and $J_{ij}<0$ correspond to favored ferromagnetic and antiferromagnetic alignment. The distances are given in {\AA}.}
\end{figure}%-------------------------------------------------------------------------------------------------------------------------------

Let us first consider the Heisenberg pair-wise exchange in the free-standing FGT monolayer
(Fig.~\ref{Heisenberg_Green}). Here, due to the lack of broken inversion symmetry, Fe3 and Fe1 are equivalent. Therefore, we end up with four different spin pairs, $\left\{ \text{Fe3-Fe3 (equivalently Fe1-Fe1), Fe2-Fe2}\right\}$ and 
%SH there was a mix up here between the Fe pairs which I fixed
$\left\{ \text{Fe2-Fe1 (equivalently Fe3-Fe2), Fe3-Fe1}\right\}$, sorted by intra- and inter-layer exchange interactions. We show in Fig.~\ref{Heisenberg_Green} the Heisenberg exchange constants
($J_{ij}$) with respect to distance ($R_{ij}$) computed by the Green's function method. All $J_{ij}$ decrease quickly with %longer 
distance, some showing an oscillatory character. The intralayer interactions 
(Fig.~\ref{Heisenberg_Green}(a,b)) favor antiferromagnetic (AFM) coupling with moderate strength of the nearest-neighbor exchange constants $J^1_{\text{Fe3-Fe3}}=-5.43$~meV and 
$J^1_{\text{Fe2-Fe2}}=-2.26$~meV while the interlayer exchange (Fig.~\ref{Heisenberg_Green}(c,d)) 
favor much stronger ferromagnetic (FM) coupling with $J^1_{\text{Fe3-Fe1}}=73.16$~meV and $J^1_{\text{Fe2-Fe1}}=18.08$~meV. In particular, the competition between FM and AFM pairs yields geometric frustration in a triangular sublattice, 
%SH
which can help
%helping 
to stabilize noncollinear spin structures \cite{leonov2015multiply,von2017enhanced,Zhang2020,Rijal2021}. Clearly, the most significant exchange coupling originates from the interaction between Fe3 and Fe1 atoms, which are on top of each other (cf.~Fig.~\ref{FGT_ge_structure}). It drops quickly to a small negative value, i.e., AFM coupling, at the second nearest distance and goes up again to a small positive value at the third nearest 
distance (Fig.~\ref{Heisenberg_Green}(d)). Similar results have been reported for FGT bulk \cite{zhu2021strain}.

\begin{figure}[t]%---------------------------------------------------------------------------------------------------------------------Fig_3
	\centering
	\includegraphics[width=1.0\linewidth]{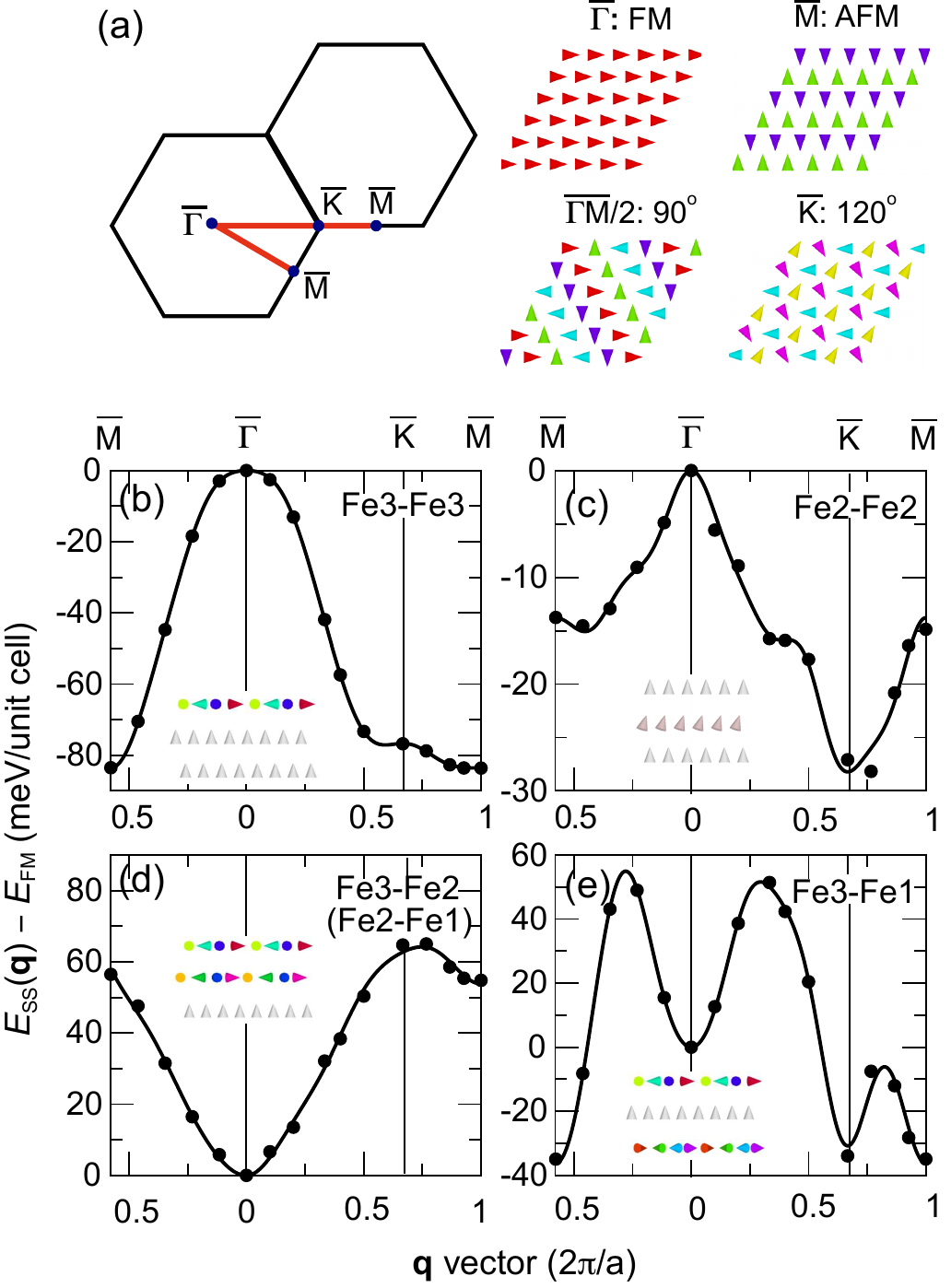}
	\caption{\label{SS} Energy dispersions of spin spirals without SOC for free-standing FGT using the FLAPW-gBT approach. (a) Sketches of the 2D hexagonal Brillouin zone and of the FM state
    ($\overline{\Gamma}$ point), the row-wise AFM state ($\overline{\text{M}}$ point), the 
    90$^{\circ}$ spin spiral (at $\overline{\Gamma {\text{M}}}$/2), and the
    N\'eel-state with 120$^{\circ}$ between adjacent spins ($\overline{\text{K}}$ point).
 %containing the high symmetry points: 
% ``$\overline{\Gamma}$", ``$\overline{\text{M}}$", ``$\overline{\Gamma {\text{M}}}$/2", ``$\overline{\text{K}}$" points which correspond to the FM state, the row-wise AFM state, the 90$^{\circ}$ spin spiral, and the N\'eel-state with 120$^{\circ}$ between adjacent spins. 
(b,c) Intralayer spin spirals for Fe3-Fe3 and Fe2-Fe2 spin pairs, respectively. (d,e) Interlayer spin spirals for Fe3-Fe1 and Fe3-Fe2 spin pairs, respectively. All energies are given relative to the energy of the FM state, $E_{\rm FM}$. Black points are from FLAPW-gBT calculations and the dispersion curve is obtained by fitting the FLAPW-gBT data to the Heisenberg model
(up to seventh NN). Insets show the spin spiral configurations in the three Fe layers of FGT
used to determine the given Fe-Fe interactions. Note, that to determine the Fe2-Fe2 exchange we used conical spin spirals in the Fe2 layer (see inset of (c)) and re-scaled the energy dispersion
(see text for details).
}
\end{figure}%-------------------------------------------------------------------------------------------------------------------------------

In order to quantitatively compare the results presented above with different DFT approaches, we have calculated the  exchange constants also by the
%repeat the calculations of the exchange constants using the 
FLAPW-gBT method. In Fig.~\ref{SS}, we present the energy dispersions $E(\vec{q})$ of ﬂat homogeneous spin spirals (per unit cell) for a free-standing FGT monolayer. The energy dispersions are calculated
%SH
in scalar-relativistic approximation, i.e.~neglecting SOC, 
along the high symmetry directions $\overline{\Gamma \text{M}}$ and $\overline{\Gamma \text{KM}}$ of the two-dimensional (2D) hexagonal Brillouin zone (BZ). 
The high-symmetry points represent special states: the $\overline{\Gamma}$ point corresponds to the FM state, the $\overline{\text{M}}$ point to the row-wise AFM state, the $\overline{\Gamma \text{M}}/2$ to the 90$^{\circ}$ spin spiral, and the $\overline{\text{K}}$ point to the N\'eel-state with 120$^{\circ}$ between adjacent spins (Fig.~\ref{SS}(a)). 
%SH move this to the caption:
%The DFT total energies
%and the fit to the Heisenberg term
%and fitted 
%(up to the seventh NN) are shown by points and solid lines, respectively. We set FM energy to zero as a reference value.

We focus first on intralayer exchange interactions. To extract only the Fe3-Fe3 exchange parameter, we rotate only Fe3 spins by fixing the Fe2 and Fe1 layers to the FM state
(inset of Fig.~\ref{SS}(b)). The lowest energy for spin spirals in the Fe3 layer is at the $\overline {\text{M}}$ point (Fig.~\ref{SS}(b)), indicating an AFM Fe3-Fe3 coupling 
in a good agreement with corresponding LCAO-Green calculations (cf.~Fig.~\ref{Heisenberg_Green}(a)). The FM state ($\overline {\Gamma}$ point) is about 83 meV/Fe atom higher in energy. 

In the case of the Fe2-Fe2 pair, the direct calculation of flat spin spiral curves propagating only in the Fe2 layer is technically unfeasible due to the complete quenching of the magnetic moment on the Fe2 atom due to symmetry (for details, see Appendix B). This shows that the magnetic moment of Fe2 is only induced by the magnetic moments of Fe1 and Fe3 in the ferromagnetic state. Therefore, we used conical spin spirals with a small cone angle of $\theta = 10^{\circ}$, i.e.~close to the FM state, and transformed the obtained energy dispersion back to a flat spin spiral: $E_{\text{flat}}(\vec{q})= E_{\text{coned}}(\vec{q}) / \sin^2{\theta} $. As shown in Fig.~\ref{SS}(c), the energy dispersion 
$E_{\text{flat}}(\vec{q})$ for the Fe2 layer becomes qualitatively different compared to the Fe3 layer: The ground state is found to be at the 
$\overline{\rm K}$ point (N\'eel state). The energy difference between FM and N\'eel state is much smaller than in the case of Fe3-Fe3, resulting in a weaker AFM Fe2-Fe2
coupling (see Table \ref{table2} in Appendix A). 

To calculate the interlayer exchange interaction, i.e.~between Fe3 and Fe2 atoms, we rotate the spins of the Fe3 and Fe2 layers simultaneously by fixing the Fe1 atom to the FM state (Fig.~\ref{SS}(d)). 
Then, we remove the intralayer contribution by calculating 
$E(\vec{q})=E_{\text{Fe3-Fe2}}(\vec{q})-E_{\text{Fe3-Fe3}}(\vec{q})-E_{\text{Fe2-Fe2}}(\vec{q})$
(not shown).
%SH: the difference curve E(q) is not shown, right?
% , as shown in Fig.~\ref{SS}(c). 
The FM state at the $\overline {\Gamma}$ point turns out to be the state of lowest energy, and the dispersion 
%SH I added E(q) to show explicitly that we are discussing this (not shown) dispersion here
$E(\vec{q})$
rises quickly for spin spirals with increasing $\vec{q}$, resulting in a strong FM coupling 
which is consistent with the LCAO-Green result (see Fig.~\ref{Heisenberg_Green}(c)). 

\begin{table}[t]
	\centering
		\caption{Comparison of the calculated Heisenberg exchange constants (in meV) for freestanding FGT using the LCAO-Green, the FLAPW-gBT, and the VASP-ML approach 
  (see Ref.~\onlinecite{xu2022assembling}). The NN intralayer (Fe3-Fe3, Fe2-Fe2) and interlayer exchange (Fe3-Fe1, Fe3-Fe2) constants are presented. The spin moments are the averaged magnetic moment which are given in $\mu_{\text{B}}$.}\label{table1}
	\scalebox{0.86}{
		\begin{tabular}{cccccc}
			\hline\hline
			\multicolumn{1}{c}{$ $} & \multicolumn{1}{c} {~~Fe3-Fe3~~} & \multicolumn{1}{c} {~~Fe2-Fe2~~} & \multicolumn{1}{c} {~~Fe3-Fe1~~} & \multicolumn{1}{c} {~~Fe3-Fe2~~} & \multicolumn{1}{c} {$M_s$/Fe} \\ 
			LCAO-Green & $-5.43$ & $-2.26$ & 73.16 & 18.08 & 1.79\\
			FLAPW-gBT  & $-10.21$  & $-2.02$ & 84.37 & 16.50 & 1.76\\
			VASP-ML \cite{xu2022assembling} & $-13.4$  & $-5.1$ & 74.1 & 39.4 &  1.66\\
			\hline
	\end{tabular}}
\end{table}	

In contrast, the energy dispersion of spin spirals propagating only in the Fe1 and Fe3 layers
%SH Fe3-Fe1 
looks very different (Fig.~\ref{SS}(e)). The energy minimum is located at the $\overline{\rm M}$
point, and two maxima are observed in the $\overline {\Gamma {\rm M}}$ and in the 
$\overline {\Gamma {\rm K}}$ directions, indicating an AFM coupling. At first glance,
this seems like a qualitative difference to the result of a strong ferromagnetic Fe3-Fe1 
coupling 
%is qualitatively different from a strong FM coupling 
observed in the LCAO-Green calculation (Fig.~\ref{Heisenberg_Green}(d)). 
However, this is due to the fact that the Fe3 and Fe1 atoms are on top of each other, and
are not distinguishable by spin spirals which propagate in-plane with respect to the Fe layers.
%$\vec{q}_x$ and $\vec{q}_y$. 
As a result, if we fit the spin spiral curve in 
Fig.~\ref{SS}(d), we obtain the shell-resolved exchange constants
between the Fe3 and Fe1 atoms, i.e.~$J_2$, $J_3$, $J_4$, etc., except for 
the nearest-neighbor term,
$J_1$, which arises from direct Fe3-Fe1 coupling. To include $J_1$ explicitly, 
we have performed an additional calculation with 
%SH
an AFM coupling between
the Fe3 and Fe1 atoms. 
%couple in AFM. 
After that, by comparing the total energies between FM and AFM, we can obtain $J_1$ 
by 
%SH carefully 
counting the number of nearest-neighbor
%SH I would move the formula to the Appendix
atoms, $E^\text{Fe3-Fe1}_{\rm AFM}-E^\text{Fe3-Fe1}_{\rm FM}=2 J_1+12 (J_2+J_3+J_4+J_6+J_7)+24 J_5$. The obtained value 
%SH
for the nearest-neighbor Fe3-Fe1 coupling is 
$J_1 = 83.64$~meV, showing a strong FM as expected. All values 
for the calculated magnetic interactions up to the seventh NN are given in Appendix A.

Our main results on the exchange constants obtained by the two computational approaches are summarized in Table \ref{table1}. For comparison, the results from a machine learning approach (denoted as VASP-ML) calculated by Xu \textit{et al.} \cite{xu2022assembling} are also included. For simplicity, we compare for all Fe-Fe pairs only the NN exchange constant. LCAO-Green, FLAPW-gBT, and VASP-ML data are in excellent qualitative agreement. Quantitatively, LCAO-Green yields exchange parameters very close to the corresponding FLAPW-gBT calculation. In particular, the Fe2-Fe2 pair agrees surprisingly well ($-2.26$ as compared to $-2.02$ meV), which indicates that the conical spin spirals used in FLAPW-gBT are very close to the Green's function method (FM state). 

\begin{figure}[t]%---------------------------------------------------------------------------------------------------------------------Fig_4
	\centering
	\includegraphics[width=0.85\linewidth]{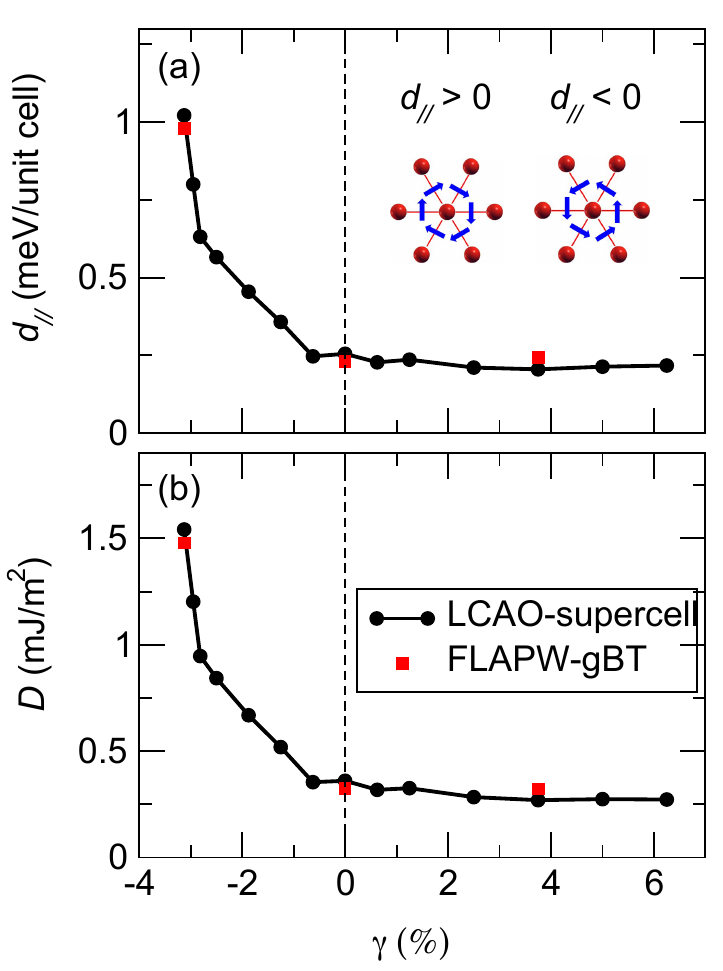}
	\caption{\label{strain} (a) Microscopic and (b) micromagnetic DMI coefficients in a monolayer FGT on germanene as a function of biaxial strain, $\gamma$, obtained by 
    the LCAO-supercell (black dots) and FLAPW-gBT (red squares) approach. The inset in (a) shows sketches of the hexagonal Fe layers, the DMI vectors, and the sign convention.}
\end{figure}%-------------------------------------------------------------------------------------------------------------------------------

However, the energies of Xu \textit{et al.} \cite{xu2022assembling} have non-negligible discrepancy. In particular, the Fe2-Fe2 and Fe3-Fe2 pairs are about 50\% higher than our FLAPW-gBT results. Reasons for this quantitative discrepancy may be the different lattice constants and different relaxations of atomic positions. Moreover, such a discrepancy might be related 
to the rather large higher-order exchange interactions (HOI) which
have been obtained for the monolayer of FGT in 
Ref.~\onlinecite{xu2022assembling}. Note, that the energy dispersions of spin spirals
include implicitly contributions from HOI which are effectively mapped to our
calculated exchange constants as discussed in detail in Ref.~\onlinecite{paul2020role}. Finally, we note that DFT+$U$ \cite{Shen2021} significantly underestimates the exchange parameters for the FGT monolayer compared to our DFT level calculations.

\subsection{FGT/Ge: Dzyaloshinskii–Moriya interaction}	

We now turn our discussion to the DMI, which plays a central %most interesting parameter
role for the emergence of non-collinear spin structures such as magnetic skyrmions.
%SH quantity studied in this work. 
The DMI originates from SOC, and it only exists in materials lacking inversion symmetry. According to Moriya’s symmetry rules \cite{Moriya1960}, since the FGT has a (001) mirror plane, $\vec{D}_{ij}$ for each pair of NN Fe atoms is perpendicular to their bonds \cite{Laref2020}. Therefore, $\vec{D}_{ij}$ can be expressed as

\begin{equation}\label{dmi_symmetry}
	\vec{D}_{ij}=d_{\parallelsum}~(\hat{\vec{u}}_{ij} \times \hat{\vec{z}})+d_{\perp}\hat{\vec{z}}
\end{equation}	

where $\hat{\vec{u}}_{ij}$ being the unit vector between sites $i$ and $j$ and $\hat{\vec{z}}$ indicating normal to the plane. 

%\textcolor{red}{In how far is this the total DMI? I would have said this is the DMI
%constant in the NN approximation.}
%The total 
The in-plane component of the DMI constant in the NN approximation, $d_{\parallelsum}$, can be directly obtained from the supercell approach \cite{YangPRL_2015} (see Eq.~(\ref{supercell})). The perpendicular component, $d_{\perp}$, does not play a significant role in 2D magnets, as reported in Ref.~\onlinecite{Liang2020,du2022spontaneous}, and is thus neglected. In the following, without specification, our calculated DMI refers to the in-plane DMI component, and $d_{\parallelsum} > 0~(d_{\parallelsum} < 0)$ denotes a CCW (CW) 
rotational sense.

For free-standing FGT, the DMI involving either Fe1 or Fe3 have opposite signs (i.e., chirality) because of the (001) mirror plane. Upon incorporating germanene in FGT, the inversion symmetry breaking at the FGT/Ge interface gives rise to an emergent DMI. Since the DMI is a key ingredient for the formation of skyrmions, methods for efficiently controlling and manipulating the DMI are essential for designing novel functional spintronics devices. We will show in the following three ways for tuning the DMI in FGT/Ge.

\begin{figure}[t]%-----------htbp------------------------------------------------------------------------------------------------------Fig_5
	\centering
	\includegraphics[width=1.0\linewidth]{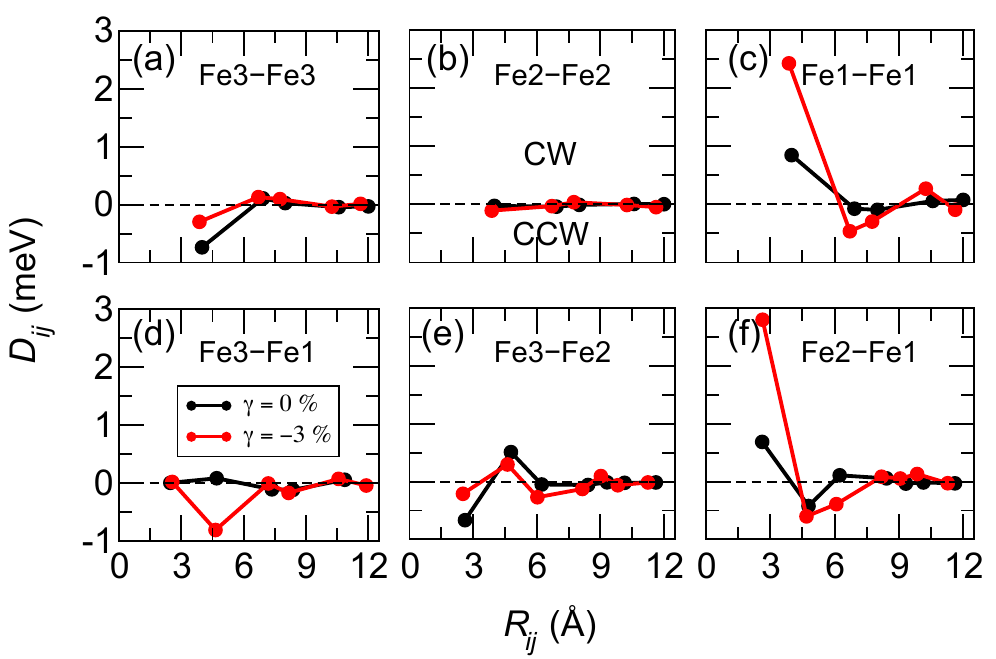}
	\caption{\label{DMI_green} The DMI parameters $D_{ij}$ between different pairs of
    Fe atoms as a function of distance for FGT/Ge at $\gamma=0\%$ (black) and $\gamma=-3\%$ (red), calculated by LCAO-Green. (a-c) intralayer and (d-f) interlayer DMI constants are presented. The positive and negative signs denote CCW and CW chirality, respectively. }
\end{figure}%-------------------------------------------------------------------------------------------------------------------------------	

\subsubsection{Strain}	

Strain engineering is one of the most commonly used methods to tune the properties of 2D layers. We study the strain-dependent DMI in FGT/Ge. The in-plane biaxial strain is defined as

\begin{equation}\label{strain_def}
	\gamma=(a-a_0)/a_0
\end{equation}
where $a$ and $a_0$ are the strained and unstrained lattice constants of the FGT, respectively. For biaxial biaxial strain ($\gamma > 0$), the in-plane lattice tends to increase while for tensile compressive strain ($\gamma < 0$) the in-plane lattice exhibits a decreasing trend.

The calculated microscopic $d_{\parallelsum}$ and micromagnetic $D$ of DMI for FGT/Ge are shown in Fig.~\ref{strain}. The DMI is evaluated quantitatively by 
%SH measuring 
calculating the self-consistent total energy of cycloidal $90^\circ$ spin spirals
%SH the systems 
with opposite rotational sense, namely using the supercell approach (i.e., LCAO-supercell). Both compressive and tensile strains are considered ranging from $-$3\% to 6.25\%. Note that 
FGT has been demonstrated to be stable under such strain via DFT phonon spectrum 
calculations~\cite{hu2020enhanced}.
%these strains have been demonstrated as stable in Ref. \onlinecite{hu2020enhanced} through the phonon spectrum calculations. 
We emphasize that the value of $a_0=4.0$~\AA~used in this work was evaluated by the GGA and
is slightly larger than the lattice constant calculated by LDA, $a_0=3.91$~\AA. If we use the latter as a reference, the largest compressive strain used in this work becomes less than $-$1\%. At equilibrium ($\gamma = 0\%$), we find a moderate DMI of about $d_{\parallelsum}=-0.25$~meV, favoring a CW spin rotation. Its corresponding micromagnetic DMI coefficient is about $|D|$ = 0.36 mJ/m$^{2}$. This value is comparable to that of
the two FGT heterostructures: FGT/In$_2$Se$_3$ \cite{Huang2022} ($\sim$0.28 mJ/m$^{2}$) and FGT/Cr$_2$Ge$_2$Te$_6$ \cite{wu2021van} ($\sim$0.31 mJ/m$^{2}$), which were demonstrated recently 
as promising vdW heterostructures to stabilize skyrmions. 

\begin{figure}[t]%-----------------htbp-------------------------------------------------------------------------------------------------Fig_6
	\centering
	\includegraphics[width=0.98\linewidth]{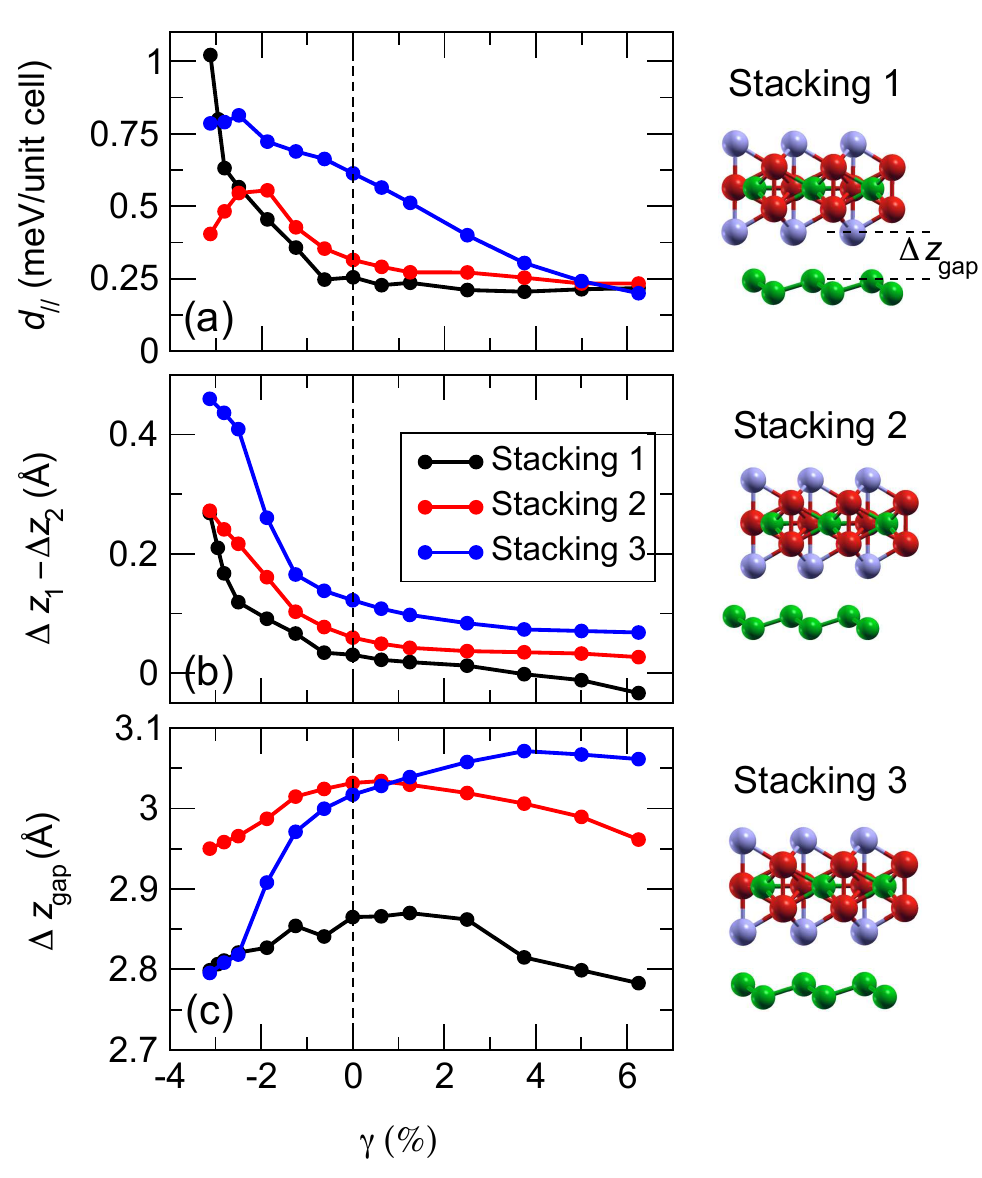}
	\caption{\label{stacking} (a) Stacking-dependent 
 %SH total 
     DMI constant in NN approximation,
     $d_\parallel$, in FGT/Ge heterostructures
     %SH I added the method here
     obtained by LCAO-supercell. The three types of stacking considered are displayed on
     the right (see text for details). 
     %SH and more text to describe the different panels
     (b) Strain dependence of $\Delta{\rm z}_1 - \Delta{\rm z}_2$ calculated from
     the Fe1-Fe2 vertical distance, $\Delta{\rm z}_1$, and the Fe2-Fe3 vertical distance, $\Delta{\rm z}_2$.
     (c) Strain dependence of the van der Waals gap $\Delta z_{\rm gap}$ (see sketch of stacking 1).}
\end{figure}%-------------------------------------------------------------------------------------------------------------------------------	

Even more interestingly, we find a significant increase in the DMI when a small compressive strain is applied (Fig.~\ref{strain}). The extracted $|D|$ is about 1.5 mJ/m$^{2}$ at $\gamma=-3\%$, which is comparable to the %previous 
value previously estimated in FM/HM thin-film systems \cite{Yang2015,woo2016observation}. Such a large DMI indicates that the FGT/Ge heterostructure
can be a promising interface for stable isolated 
%SH I would cite our prediction of skyrmions in FGT/Ge here as well
skyrmions \cite{Dongzhe2022_fgt}.
A more detailed analysis shows that the enhancement of the DMI in FGT/Ge is mainly medicated by the geometrical change due to the buckled structure of germanene as shown 
in Sec.~\ref{sec_stacking}. 

The results presented above are reproduced by FLAPW-gBT if we calculate the effective DMI from spin spiral calculations including 
SOC (red squares in Fig.~\ref{strain}).
%which is obtained by an effective model with values as fitted from spin spiral calculations, including SOC (red squares). 
Here, the effective DMI constant is obtained from fits in the region of low $|\vec{q}|$ around the 
FM state ($\overline {\Gamma}$ point) where the energy contribution due to SOC varies linearly
with $|\vec{q}|$. Interestingly, the DMI predicted by the LCAO-supercell method is in excellent quantitative agreement with the one calculated by FLAPW-gBT, showing that the reported results are robust against different approaches used. This indicates a possibility of strain control of magnetic skyrmions in these systems. In contrast, a tensile strain ($\gamma>0$) has almost no effect on the DMI (Fig.~\ref{strain}).

To gain further insight into the local decomposition of the DMI, we show in Fig.~\ref{DMI_green} the calculated DMI using the LCAO-Green approach for six possible Fe pairs in FGT/Ge as a function of distance with and without strain. Evidently, the external 
compressive strain ($\gamma=-3$~\%)
has a significant effect on the DMI of the FGT monolayer, particularly for pairs connected to the interface Fe1 atom. 
%SH (Figs.~\ref{DMI_green}(c,f)). 
When the strain is applied, a strong enhancement of the DMI favoring a CW (i.e., 
positive sign) rotational sense is clearly observed for Fe1-Fe1 and Fe2-Fe1 pairs (Fig.~\ref{DMI_green}c,f), indicating a strong modification of electronic and magnetic properties at the interface. This is also reflected by the projected density of states in FGT/Ge (see Fig.~\ref{pdos} in Appendix C). 

%\textcolor{red}{I think the signs of the DMI in this paragraph should be opposite. Please check.}
Quantitatively, the
nearest-neighbor DMI changes from  0.88 (0.72) meV to 2.46 (2.84) meV for the 
Fe1-Fe1 (Fe2-Fe1) pair. A similar behavior is also observed for the second-nearest neighbor DMI in Fe3-Fe1 (Fig.~\ref{DMI_green}d). The DMI changes from 0.1 meV to $-$0.80~meV, accompanying a sign change. On the other hand, Fe3-Fe3 and Fe3-Fe2 (Figs.~\ref{DMI_green}(a,e) are much less affected by the strain. Moreover, due to symmetry, the DMI in Fe2-Fe2 (Fig.~\ref{DMI_green}(b)) and the nearest neighbor Fe3-Fe1 (on top of each other) are almost quenched. These results are also 
%SH reproduced 
in agreement with those obtained by the {\tt TB2J} code \cite{he2021tb2j} which uses a similar methodology.

	\begin{table*}[t]
		\centering
		\scalebox{0.97}{
			\begin{tabular}{cccccccccccccccccc}
				\hline\hline
				\multicolumn{1}{c}{$ $} & \multicolumn{1}{c} {~~$J_{1}$~~} & \multicolumn{1}{c} {~~~$J_{2}$~~~} & \multicolumn{1}{c} {~~~$J_{3}$~~~} & \multicolumn{1}{c} {~~~$J_{4}$~~~} & \multicolumn{1}{c} {~~~$J_{5}$~~~} & \multicolumn{1}{c} {~~~$J_{6}$~~~} & \multicolumn{1}{c} {~~~$J_{7}$~~~} & \multicolumn{1}{c} {~~~$J_{8}$~~~} & \multicolumn{1}{c} {~~~$D_{1}$~~~} & \multicolumn{1}{c} {~~~$D_{2}$~~~} & \multicolumn{1}{c} {~~~$D_{3}$~~~} & \multicolumn{1}{c} {~~~$D_{4}$~~~} & \multicolumn{1}{c} {~~~$D_{5}$~~~} & \multicolumn{1}{c} {~~~$D_{6}$~~~} & \multicolumn{1}{c} {~~~$D_{7}$~~~} \\ 
				$+0.5$ V/\AA~ & 23.73 & $0.16$ & $-1.61$ & 0.48 & 1.12& $-0.28$ & 0.00 & $-0.15$ & 0.31 & 0.09 & $-0.01$ & $-0.01$ & $-0.01$ & 0.00 & 0.00  \\
				$~~$0.0 V/\AA~ & 22.87 & $-0.21$ & $-1.78$ & 0.43 & 1.25& $-0.31$ & 0.00 & $-0.15$ & 0.23 & 0.10 & 0.00 & 0.00 & 0.00 & 0.00 & 0.00  \\
				$-0.5$ V/\AA~  & 22.35  & $-0.62$ & $-1.90$ & 0.38 & 1.30 & $-0.37$ & 0.06 & $-0.17$ & 0.14 & 0.08 & $-0.01$ & $-0.01$ & $-0.01$ & 0.00 & 0.00 \\
				\hline
		\end{tabular}}
		\caption{Shell-resolved Heisenberg exchange constants ($J_i$) and DMI constants ($D_i$) obtained by fitting the energy contribution to spin spirals without and with
SOC from DFT calculations using the FLAPW method as presented in Fig.~\ref{efield_J} and Fig.~\ref{efield} for three electric field values. A positive (negative) sign of $D_i$ denotes a preference of CW (CCW) rotating cycloidal spin spirals. All values are given in meV/unit cell.} \label{table_efield}
	\end{table*}		

\subsubsection{Stacking} \label{sec_stacking}	

To study the effect of stacking order on the DMI of the FGT/Ge heterostructure, three fully optimized representative stacking geometries
%patterns 
are considered here (see sketches in Fig.~\ref{stacking}): 
(i) stacking 1 (the most favorable stacking
geometry), in which the Te atom is right above the center of the hexagonal ring of germanene, 
(ii) stacking 2, in which the Ge atom of FGT is located above the center of the hexagonal ring of germanene, and (iii) stacking 3, in which the Fe1 and Fe3 atoms of FGT are placed directly above the center of the hexagonal ring of germanene.

Fig.~\ref{stacking}(a) shows the variation of the DMI strength $d_{\parallelsum}$ with respect to strain for these three stacking geometries. We observe a similar behavior for stackings 1 and 2 
for the range of $\gamma$ between about $-$2\% to 6\%, namely, a clear enhancement of DMI at
compressive strain ($\gamma<0$)
and a nearly constant value for a tensile strain ($\gamma>0$).
%is observed when a compress strain is applied. 
Interestingly, when $\gamma < -2\%$, the DMI increases for stacking 1 while it decreases for stacking 2. 
%For stacking 3, the DMI is already much increased at the equilibrium configuration,
%i.e.~$\gamma=0$, and shows a nearly linear dependence on strain between $-$2\% to 5\%.

To understand the origin of the variation of the DMI, we plot in Fig.~\ref{stacking}(b-c) the geometrical change by strain. As a negative strain is applied, $\Delta{\rm z}_1 - \Delta{\rm z}_2$ increases rapidly for both stackings 1 and 2, where $\Delta{\rm z}_1$ and $\Delta{\rm z}_2$ denote the vertical distances between Fe1-Fe2 and Fe2-Fe3. This leads to an increase in the degree of structural asymmetry, resulting in an enhancement of the DMI. On the other hand, the vdW gap induced by hybridization is decreased by strain and is more pronounced for stacking 2 than for stacking 1. As analyzed in detail in Ref.~\cite{Dongzhe2022_fgt}, the hybridization effect for FGT/Ge decreases the DMI. This explains why the DMI for stacking 2 is smaller than 
for stacking 1 for $\gamma < -2\%$. 

\begin{figure}[tp]%-----------------htbp-------------------------------------------------------------------------------------------------Fig_7_0
	\centering
	\includegraphics[width=0.95\linewidth]{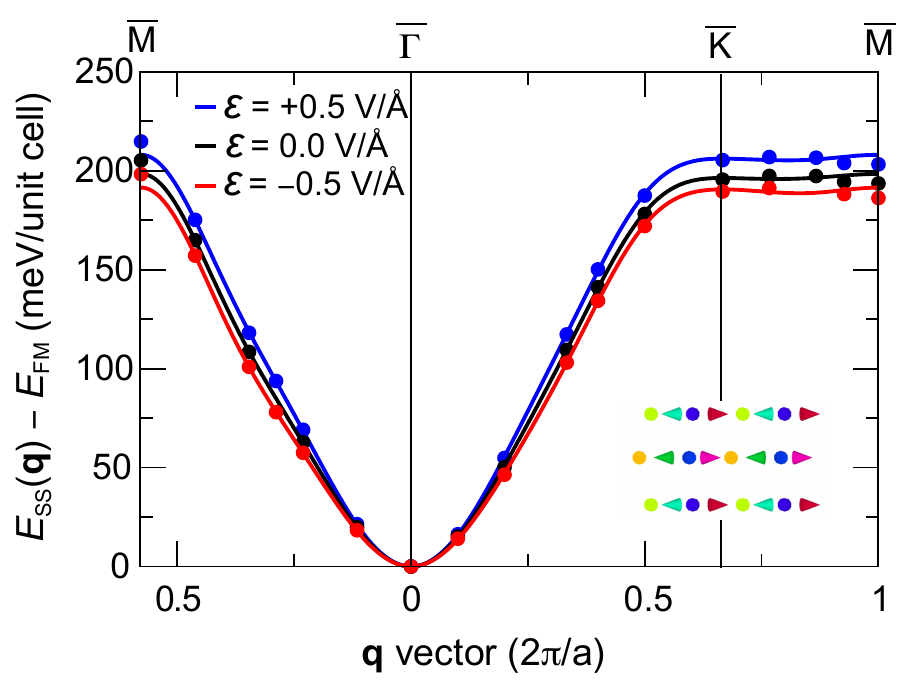}
	\caption{\label{efield_J} Energy dispersions of spin spirals propagating in all Fe layers 
    (see inset)
    of FGT/Ge in scalar-relativistic approximation, i.e.
  %SH curves 
 without SOC, calculated via the FLAPW method along two high symmetry directions ($\overline{\Gamma{\text{K}}{\text{M}}}$ and $\overline{\Gamma \text{M}}$) at
 electric fields of
 ${\cal E} = +0.5$~V/\AA~(blue), ${\cal E} = 0.0$~V/\AA~(black), and ${\cal E} = -0.5$~ V/\AA~(red). The filled circles represent DFT total energies and the solid lines are fits to the Heisenberg model 
(see Table \ref{table_efield} for exchange constants).
 }
\end{figure}%-------------------------------------------------------------------------------------------------------------------------------	

For stacking 3, we observe the most pronounced geometrical changes, the variation of $\Delta{\rm z}_1 - \Delta{\rm z}_2$ and $\Delta_{\text{gap}}$ are almost two times and three times larger than for stacking 1, respectively. As a result, even at $\gamma = 0\%$ the DMI for stacking 3 is about $-0.63$ meV, which is 2.5 times higher than the corresponding DMI in stacking 1. However, only a slight increase of the DMI up to about $-0.78$ meV is observed with compressive strain. Herein, the increase of $d_{\parallelsum}$ is mainly due to the interplay between the increase of $\Delta{\rm z}_1 - \Delta{\rm z}_2$ and the decrease of $\Delta_{\text{gap}}$, which have an opposite effect on the DMI. By applying a tensile strain to the system, the DMI decreases nearly linearly for stacking 3.

\begin{figure}[tp]%-----------------htbp-------------------------------------------------------------------------------------------------Fig_7
	\centering
	\includegraphics[width=0.95\linewidth]{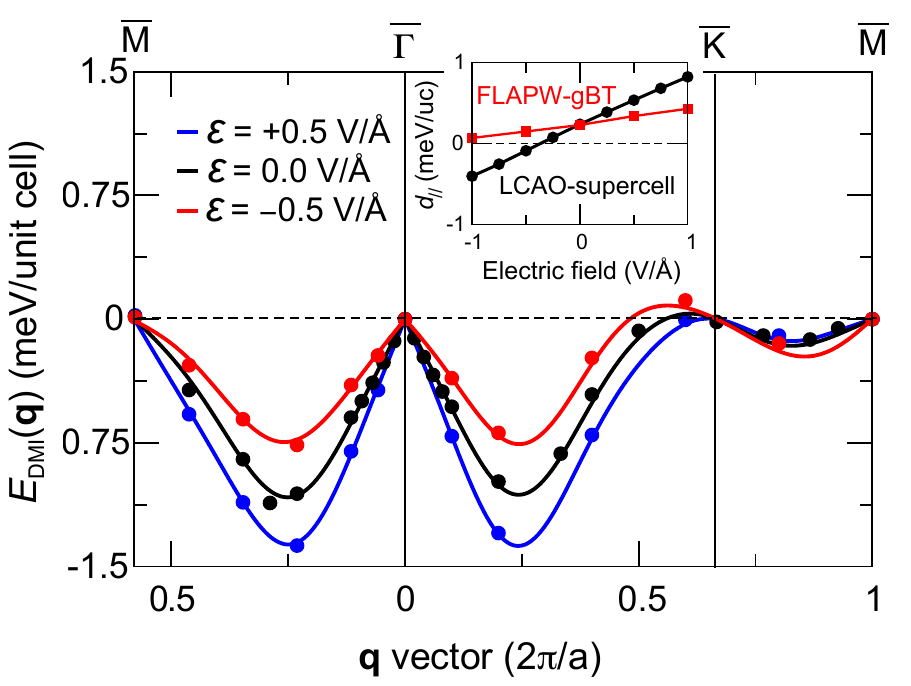}
	\caption{\label{efield} SOC induced energy contribution $E_{\text{DMI}}(\V{q})$ to the
     dispersion of flat cycloidal spin spirals propagating in all Fe layers of FGT/Ge
     calculated via the FLAPW method
     along the high symmetry directions ($\overline{\Gamma{\text{K}}{\text{M}}}$ and $\overline{\Gamma \text{M}}$) at ${\cal E} = +0.5$~V/\AA~(blue), ${\cal E} = 0.0$~V/\AA~(black), and ${\cal E} = -0.5$~ V/\AA~(red). The filled circles represent DFT data and the solid lines are fits to the DMI term of the 
     %extended Heisenberg 
     atomistic spin model. The inset shows the 
     %SH total 
     DMI constant in NN approximation for the FGT/Ge 
    heterostructure as a function of the perpendicular electric field, calculated by the
    LCAO-supercell (black) and the value of the NN DMI constant obtained
    in the FLAPW-gBT (red) approach (for all values see Table \ref{table_efield}).
    }
\end{figure}%-------------------------------------------------------------------------------------------------------------------------------	

\subsubsection{Electric field}	

The FLAPW method in film geometry as implemented in the {\tt FLEUR} code is used to include the effect of an external electric field as described in Ref.~\onlinecite{Weinert2009}. An out-of-plane electric field is defined by adding a charge plate in the vacuum and adding the same amount of opposite charge to FGT/Ge (cf.~Fig.~\ref{FGT_ge_structure})
to maintain charge neutrality \cite{Weinert2009,Oba2015,paul2022electric}.
We chose electric field values of ${\cal E}=\pm0.5$ V/\AA, which can be applied in scanning tunneling microscopy experiments \cite{hsu2017electric}
and allow to write and delete isolated magnetic skyrmions in ultrathin transition-metal films as demonstrated experimentally \cite{hsu2017electric}
and based on atomistic spin simulations with DFT parameters
\cite{paul2022electric,Goerzen2022}. Note, that we use the same sign convention for the electric field as in Refs.~\onlinecite{paul2022electric,Goerzen2022}.

%SH I added a few more details here
We show in Fig.~\ref{efield_J} the energy dispersion E(\V{q}) of homogeneous flat spin spirals propagating in all Fe layers of FGT/Ge in the scalar-relativistic approximation, i.e.~without SOC. Upon including a negative
%SH positive 
electric field with a strength of 
${\cal E}=-0.5$ V/\AA, the energy rises more slowly at the $\overline{\Gamma}$ 
point (FM state) than for ${\cal E}=0$
and the energy difference with respect to the AFM state 
($\overline{\rm M}$ point) and N\'eel state ($\overline{\rm K}$ point) decreases.
For a positive
%SH negative 
electric field of ${\cal E}=+0.5$ V/\AA, we observe the opposite
effect, i.e.~a faster rise and larger energy differences at the high symmetry
points. This field effect is similar to that observed for Fe monolayers
on transition-metal surfaces \cite{paul2022electric}.

By fitting the energy dispersions with and without an applied electric field, we obtain
the field effect on the exchange constants (see Table \ref{table_efield}). Note that we treat three Fe atoms as a whole in this spin model, i.e.~without explicitly considering the exchange interactions between different Fe pairs. In other words, the definition of $J_{i}$ is different from the one defined in Section \ref{free-layer}. Here, the exchange interactions between different Fe pairs are included in an averaged way. We find a nearly linear decrease of the nearest-neighbor exchange constant, $J_1$, by about 6\% upon applying an electric field of 1~V/{\AA}. The exchange constants beyond nearest neighbors are also significantly influenced by the electric field which shows that the 
exchange frustration increases for a positive electric field. The effect of
the electric field on the exchange interaction can be explained based on its
spin-dependent screening at the surface as shown previously
for ultrathin $3d$ transition-metal films \cite{Oba2015,paul2022electric}.

Fig.~\ref{efield} shows the SOC induced DMI contributions to the energy dispersion of cycloidal spin spirals in FGT/Ge, $E_{\text{DMI}}({\V{q}})$, under electric field. When an electric field is applied, $E_{\text{DMI}}({\V{q}})$ displays the same trend as in zero field, i.e., it favors cycloidal spirals with a CW rotational sense, but an electric-field induced modification in $E_{\text{DMI}}({\V{q}})$ is clearly seen. The 
out-of-plane electric field breaks the inversion symmetry and leads to a drastic change
of the nearest-neighbor DMI constant, $D_1$, which increases and decreases by approximately 50\% for ${\cal E}=+0.5$ V/\AA~and ${\cal E}=-0.5$ V/\AA, respectively (see Table \ref{table_efield}). Additionally, we find that $D_2$ remains almost the same. 

In general, when an electric field is applied from negative to positive, the DMI favors less and less a CW rotational sense. This is also reproduced by the LCAO-supercell approach (see inset of Fig.~\ref{efield}) in which we even find a change of sign of the DMI constant at 
${\cal E}<-0.37$~V/{\AA}. However, quantitatively, we note that LCAO-supercell overestimates the DMI compared to FLAPW-gBT under an electric field. We attribute this quantitative discrepancy to the different implementations of the electric field  and basis sets used in the FLAPW-gBT and LCAO-supercell methods. 

%\textcolor{red}{I started a new paragraph here, since the following paragraph
%discusses another issue/point.}
Note, that the electric-field effect on the exchange interaction is opposite to that on the DMI  
with respect to the formation of non-collinear spin states. For 
%SH2: I think the efield is now defined in the opposite sense
${\cal E}<0$, 
the energy dispersion without SOC (Fig.~\ref{efield_J}) rises less quickly and the exchange
frustration increases which is favorable for the emergence of spin structures such as
skyrmions. However, the energy contribution due to DMI (Fig.~\ref{efield}) shows a
shallower energy minimum for ${\cal E}<0$ and the nearest neighbor DMI constant 
drops accordingly (cf.~Table \ref{table_efield}). This
shows that both the effect on the exchange and on the DMI needs to be taken into account
in order to predict electric-field assisted formation of non-collinear spin structures
in line with previous theoretical studies \cite{paul2022electric,Goerzen2022}. In contrast, the electric-field effect on the magnetocrystalline 
anisotropy energy is small (see next section).

\begin{figure}[t]%-----------------htbp-------------------------------------------------------------------------------------------------Fig_8
	\centering
	\includegraphics[width=1.0\linewidth]{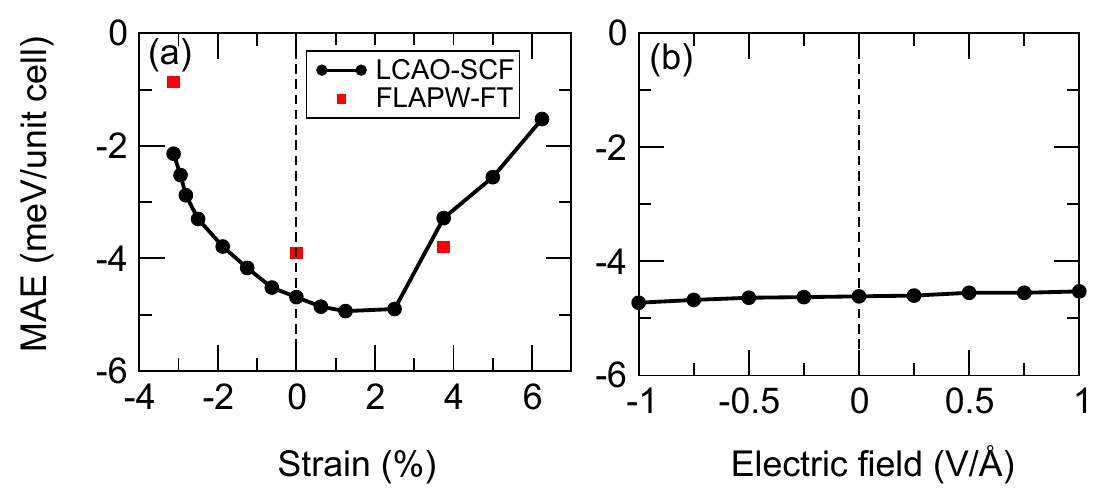}
	\caption{\label{mae} (a) Strain and (b) electric-field dependent MAE in the FGT/Ge heterostructure obtained by the LCAO-SCF (black dotted) and the FLAPW-FT (red square) approach.}
\end{figure}%-------------------------------------------------------------------------------------------------------------------------------	

\subsection{FGT/Ge: Magnetocrystalline anisotropy energy}	

We define the MAE as the difference in total energies between a configuration in which the
magnetization in the ferromagnetic state
is in-plane ($\parallel$) and out-of-plane ($\perp$)
with respect to the FGT monolayer
\begin{equation}\label{mca}
\text{MAE}=E_{\perp}-E_{\parallel}
\end{equation}	

In Fig.~\ref{mae}, we plot the MAE in the FGT/Ge heterostructure as a function of strain
and electric field. We calculate the MAE using two approaches: (i) The MAE is defined as the energy difference between self-consistently converged total energies including SOC, namely LCAO-SCF and (ii) the MAE is taken as the band energy difference (also known as force theorem) obtained after a one-step diagonalization of the full Hamiltonian including SOC, starting from the well converged self-consistent scalar relativistic (without SOC) density/potential, namely FLAPW-FT. 

We find that FGT/Ge has a strong perpendicular MAE of more than 4.62 meV (about 1.5 meV/Fe). Interestingly, when a mechanical strain is applied, the MAE can be significantly reduced to 2.08 meV at $\gamma=-3\%$ and 1.44 meV at $\gamma=6.1\%$, respectively. In general, the two DFT approaches yield a good qualitative agreement concerning the change of the MAE with strain. From the local decomposition of MAE evaluated by the grand-canonical formulation \cite{Dongzhe2013,Dongzhe2014}, we find that Fe1 and Fe3 favor favor out-of-plane anisotropy while the Fe2 layer favors an in-plane direction of the moments. In contrast, we find that an applied electric field has a much smaller effect on the MAE of FGT/Ge (Fig.~\ref{mae}(b)).

\section{Conclusion} 
\label{concl}

In summary, Fe$_3$GeTe$_2$/germanene has been investigated as a representative 2D vdW 
magnetic heterostructure using three current state-of-the-art approaches to map the \textit{ab initio} DFT calculations to an atomistic spin model:
%SH solutions to an extended Heisenberg model: 
(i) The Green's function approach performing infinitesimal rotations, (ii) the spin spiral method employing the generalized Bloch theorem for various $\V{q}$ vectors, and (iii) the supercell approach based on the chirality-dependent total energy difference. We obtain good qualitative agreement for the Heisenberg exchange and Dzyaloshinskii-Moriya interaction in FGT/Ge using these three different approaches. We obtain almost quantitative agreement for the DMI between methods (ii) and (iii), indicating that the nearest-neighbor approximation is valid in the FGT/Ge heterostructure. Furthermore, we have studied the electronic and magnetic ground states of the FGT/Ge heterostructure under biaxial mechanical strain, stacking, and a perpendicular electric field. We have shown that the strength of the DMI can be significantly modified via strain and stacking order, tracing its origin to the geometrical change and hybridization effect. In particular, when a small compressive strain is applied, the DMI is strongly enhanced while the MAE, in contrast, is significantly decreased, which allow the possibility of nanoscale skyrmions at a low magnetic field~\cite{Dongzhe2022_fgt}. On the other hand, an electric field changes the DMI and the exchange constants almost linearly. 
If we apply a large enough electric field, we also expect a change of the sign of the DMI,
i.e.~a reversed rotational sense of the favored non-collinear spin structures. The electric field effect on the magnetocrystalline anisotropy is small for the 
FGT/Ge heterostructure.

%%%%%%%%%%%%%%%%% Acknowledgement %%%%%%%%%%%%%%%%%%%%%%%
\section*{Acknowledgments}

This study has been supported through the ANR Grant No. ANR-22-CE24-0019. This study has been (partially) supported through the grant NanoX no.~ANR-17-EURE-0009 in the framework of the ``Programme des Investissements d’Avenir". S.~Ha, T.~D.~and S.~He.~gratefully acknowledge financial support from the Deutsche Forschungsgemeinschaft (DFG, German Research Foundation) through SPP2137 ``Skyrmionics" (project no.~462602351). This work was performed using HPC resources from CALMIP (Grant 2023-[P21008]). D.~Li thanks F. Nickel for valuable discussions.

%%%%%%%%%%%%%%%%% Acknowledgement %%%%%%%%%%%%%%%%%%%%%%%

\section*{Appendix A: Fitted Heisenberg exchange parameters for free-standing FGT}

In Table \ref{table2}, we present the fitted parameters up to the seventh nearest neighbors for free-standing FGT. Note that, in the main text, we only benchmarked the NN exchange parameter.

\begin{figure}[tp]%-------------------htbp-----------------------------------------------------------------------------------------------Sup1
	\centering
	\includegraphics[width=1.0\linewidth]{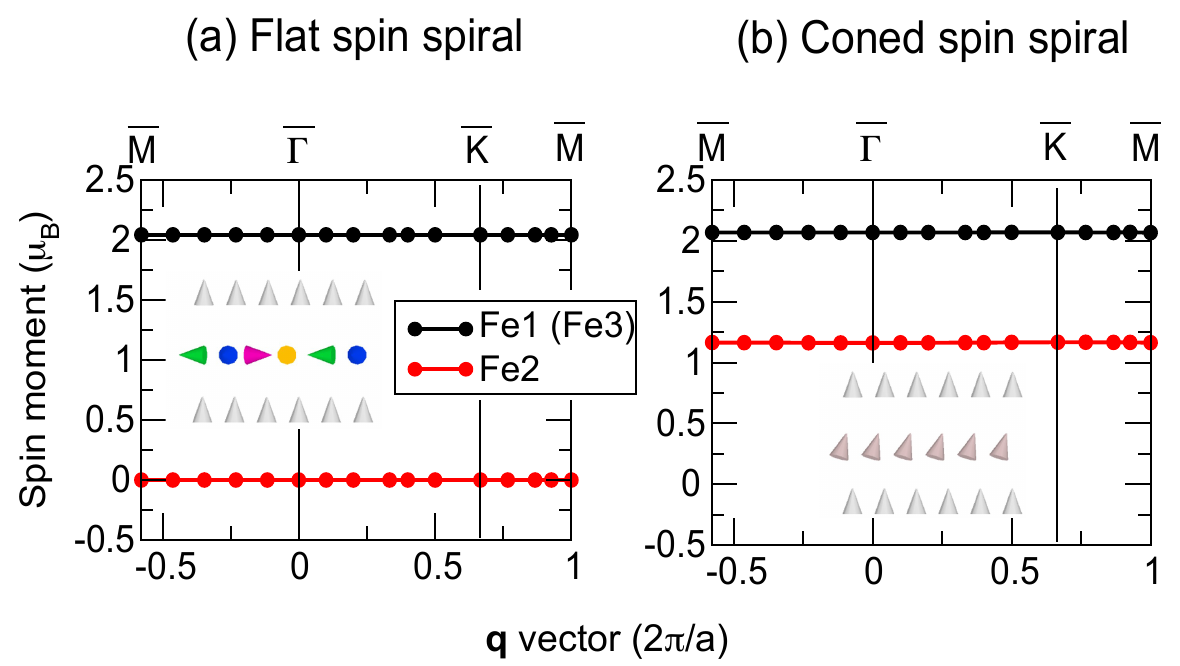}
	\caption{\label{coned_flat} Magnetic moments of Fe1 (Fe3) and Fe2 atom in a freestanding FGT monolayer as a function of the $\vec{q}$-vector for flat ($\theta = \pi/2$) and conical (with a small cone angle of $\theta = \pi/20$) spin spirals calculated via the FLAPW method.}
\end{figure}%--------------------------------------------------------------------------------

\begin{table}[h!]
\centering
\caption{Shell-resolved Heisenberg exchange constants, $J_i$, for $i$-th neighbors for the different 
Fe-Fe pairs in FGT fitted from the spin spiral DFT calculations via the FLAPW method presented in Fig.~\ref{Heisenberg_Green}. All values are given in meV.}
\scalebox{0.78}{
	\begin{tabular}{cccccccc}
			\hline\hline
			\multicolumn{1}{c}{$ $} & \multicolumn{1}{c} {~~~~$J_1$~~~~} & \multicolumn{1}{c} {~~~~$J_2$~~~~} & \multicolumn{1}{c} {~~~~$J_3$~~~~} & \multicolumn{1}{c} {~~~~$J_4$~~~~} & \multicolumn{1}{c} {~~~~$J_5$~~~~} & \multicolumn{1}{c} {~~~~$J_6$~~~~} & \multicolumn{1}{c} {~~~~$J_7$~~~~}\\ 
			Fe3-Fe3 (Fe1-Fe1) & $-$10.42 & $-$0.30 & 1.28 & 0.19 & $-$0.28 & 0.03 & 0.04 \\
			Fe2-Fe2  & $-$2.02  & 0.50 & $-$0.53 & $-$0.20 & $-$0.01 & 0.03 & $-$0.38 \\
			Fe3-Fe2 (Fe2-Fe1) & 16.50  & 5.42 & $-$1.23 & $-$0.82 &  1.01 & 1.11 & 0.50 \\
			Fe3-Fe1 & 83.47  & $-$6.40 & 4.13 & 6.86 &  $-$1.08 & 1.09 & 1.25 \\
			\hline
			\end{tabular}}
		 \label{table2}
		\end{table}

\section*{Appendix B: Quenching of magnetic moment}

Spin spirals can be characterized by their reciprocal spin spiral vector $\vec{q}$, which determines the propagation direction of the spiral. For a rotation axis along the $z$ direction, their magnetization is defined as

\begin{equation}\label{def_spinspiral}
\vec{m}_i =
\begin{pmatrix}
\cos(\vec{q} \cdot \vec{R}_i) \sin\theta\\
\sin(\vec{q} \cdot \vec{R}_i) \sin\theta \\
\cos\theta
\end{pmatrix}
\end{equation}

where $\vec{R}_i$ is the position of site $i$ and $\theta$ is the cone angle. For the special value $\theta = \pi /2$, we obtain flat spin spirals.

%-----------------------------------------------	
	\begin{figure}[b]%-----------------htbp-----------------------------------------------------------------------------------------------Sup3
	\centering
	\includegraphics[width=1.0\linewidth]{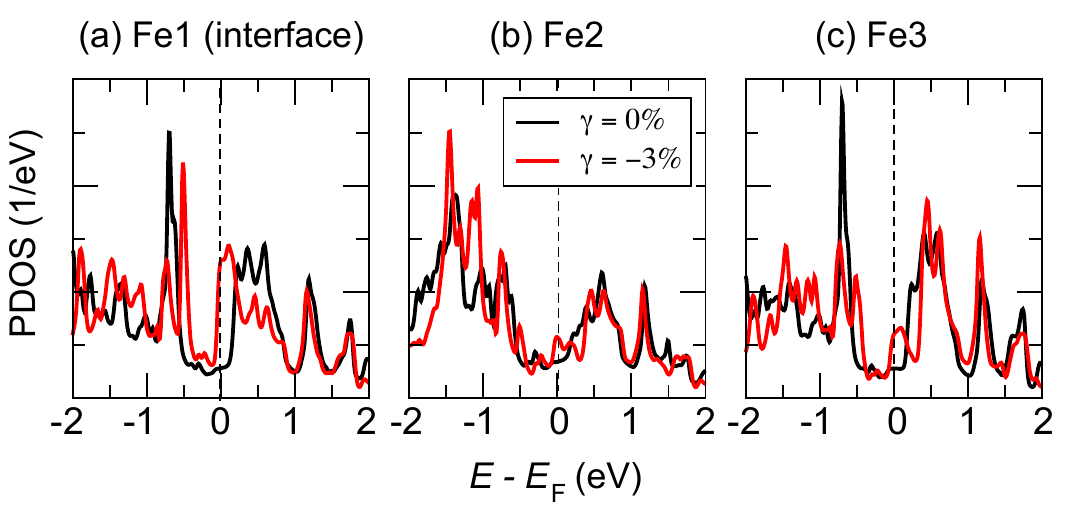}
	\caption{\label{pdos}  Calculated spin-averaged PDOS on (a) Fe1, (b) Fe2, and (c) Fe3 atoms in FGT/Ge bilayer in equilibrium, i.e.~$\gamma=0\%$ (black), and under a
    compressive strain of $\gamma=-3\%$ (red) obtain by the LCAO-Green approach. 
    The Fermi energy is set to zero.}
\end{figure}%-------------------------------------------------------------------------------------------------------------------------------

We show in Fig.~\ref{coned_flat} the variation of magnetic moments along $\vec{q}$ vectors for (a) flat and (b) conical spin spirals for Fe2-Fe2 spin spirals in a freestanding FGT monolayer, 
where we rotate only the spins of the Fe2 atoms by fixing other two layers to the FM state. For flat spin spirals (Fig.~\ref{coned_flat}(a)), only the Fe3 and Fe1 layers possess a magnetic moment, while the moments of the Fe2 atom vanish due to symmetry. On the other hand, we obtain
the expected magnetic moments (close to the FM state) for all three atoms. 

\section*{Appendix C: The PDOS in FGT/Ge without and with strain}

To further show that the mechanism of strong enhancement of DMI originates mainly from
the Fe1 atom related pairs (see Fig.~\ref{DMI_green}), the PDOS of Fe atoms 
in FGT/Ge under a strain of $\gamma=-$3\% and 0\% are analyzed in Fig.~\ref{pdos}. The largest difference between the two cases
%$\gamma=0\%$ and $\gamma = -3\%$ 
originates from the Fe1 atom, which is located at the interface. 

\section*{Appendix D: Comparison of {\tt FLEUR} and {\tt QATK}}

\begin{figure}[h]%-----------------htbp-----------------------------------------------------------------------------------------------Sup4
	\centering
	\includegraphics[width=1.0\linewidth]{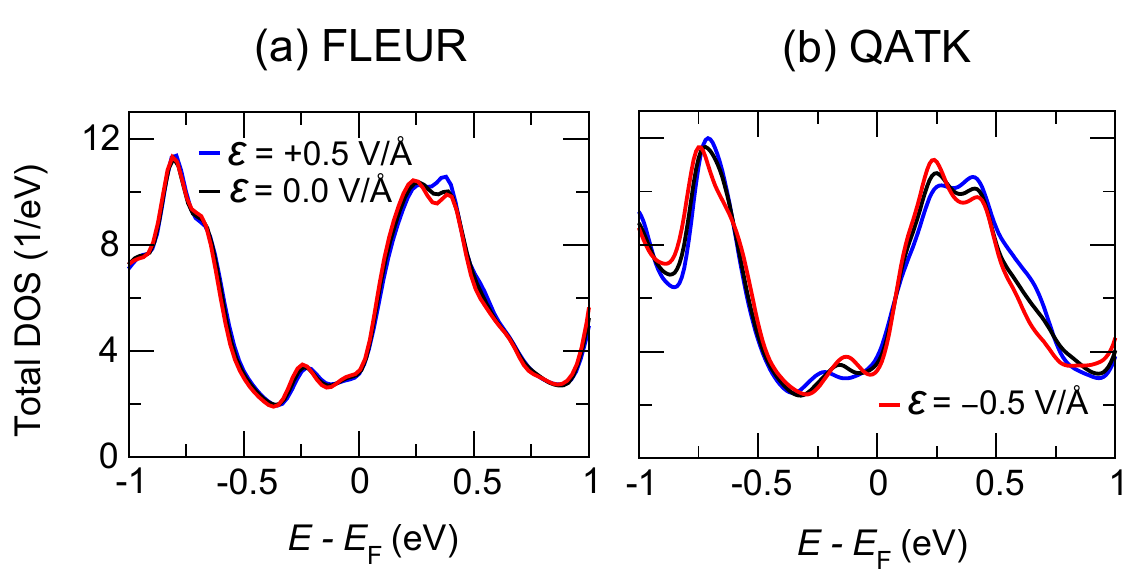}
	\caption{\label{pdos_com}  Total DOS of FGT/Ge obtained via (a) the {\tt FLEUR} code and 
 (b) {\tt QATK} 
 %SH for the total DOS of FGT/Ge 
 at electric fields of ${\cal E} = +0.5$~V/\AA~(blue), ${\cal E} = 0.0$~V/\AA~(black), and ${\cal E} = -0.5$~ V/\AA~(red).}
\end{figure}

We plot in Fig.~\ref{pdos_com} the total DOS for FGT/Ge calculated by the {\tt FLEUR} 
code (left) and by the {\tt QATK} code (right) at different electric fields. We note that the effect of the electric field on the PDOS is much more pronounced in {\tt QATK} 
%is much larger 
than in {\tt FLEUR}, leading to the quantitative difference in the variation of the
DMI with electric field observed in Fig.~\ref{efield} (inset).

	%------------ 

\bibliography{References}
	
\end{document}